\documentclass[aps,pra,twocolumn]{revtex4-1}
\usepackage[utf8]{inputenc}
\usepackage[dvips]{graphicx}
\usepackage{epsfig}

\usepackage[margin=2cm,head=0.5cm]{geometry}
\usepackage{bm}
\usepackage{amsmath}
\usepackage{amssymb}
\usepackage{latexsym}
\usepackage{amsfonts}
\usepackage{epsfig}
\usepackage{color}
\usepackage[linktocpage, colorlinks=true ,linkcolor=blue, citecolor=blue]{hyperref}
\usepackage[all]{hypcap}
\usepackage{braket}
\usepackage{dsfont}

\newcommand{\expval}[1]{\left< #1 \right>}

\newcommand{\nn}{\nonumber\\}

\newcommand{\f}[1]{\mbox{\boldmath$#1$}}

\newcommand{\ord}{{\cal O}}
\newcommand{\trace}[1]{{\rm Tr}\left\{ #1 \right\}}
\newcommand{\ptrace}[2]{{\rm Tr_{#1}}\left\{ #2 \right\}}
\newcommand{\traceB}[1]{{\rm Tr_B}\left\{ #1 \right\}}

\newcommand{\abs}[1]{{\left| #1 \right|}}

\newcommand{\sgn}{{\rm sgn}}
\newcommand{\sinc}{{\rm sinc}}

\newcommand{\ii}{\mathrm{i}}  

\begin{document}
\title{Coarse-graining master equation for periodically driven systems}
\author{Ronja Hotz$^1$}
\author{Gernot Schaller$^{1,2}$}
\email{g.schaller@hzdr.de}
\affiliation{$^1$ Institut für Theoretische Physik, Technische Universit\"at Berlin, Hardenbergstr. 36, 10623 Berlin, Germany}
\affiliation{$^2$ Helmholtz-Zentrum Dresden-Rossendorf, Bautzner Landstraße 400, 01328 Dresden, Germany}
\date{\today}

\begin{abstract}
We analyze Lindblad-Gorini-Kossakowski-Sudarshan-type generators for selected periodically driven open quantum systems.
All these generators can be obtained by temporal coarse-graining procedures, and we compare different coarse-graining schemes.
Similar as for undriven systems, we find that a dynamically adapted coarse-graining time, effectively yielding non-Markovian dynamics by interpolating
through a series of different but invididually Markovian solutions, gives the best results among the different coarse-graining schemes, albeit at highest computational cost.
\end{abstract}

\maketitle

\section{Introduction}

While the propagation of undriven closed quantum systems via Schr\"odingers equation may be challenging for many degrees of freedom, it can be calculated with standard methods. 
This changes when the Hamiltonian is subject to an external driving, when time-ordering becomes relevant.
Even for the simple case of periodic driving, where Floquet theory~\cite{Floquet1883,PhysRev.138.B979} applies, the exact calculation of the unitary propagator may be notoriously difficult~\cite{GRIFONI1998229,eckardt2015a,bukov2015a}. 
In addition, it is then often hidden that the time-dependent variation of external parameters requires that the system is coupled in some way to an outside world, which to be complete would actually require to model the system as open.
While open quantum systems in absence of driving are well-studied and interesting-subject on their own~\cite{weiss1993,book_breuer,schlosshauer2007}, the joint discussion of the effects of system-reservoir coupling and periodic driving is challenging.

However, with in particular periodic types of driving being experimentally quite feasible with current standards~\cite{CHU861,book_rice_zhao,
Aidelsburger_2013,Atala_2014,Jotzu_2014}, the derivation of proper dissipators for periodically driven open systems is of great concern.
In principle, it is formally possible to perform the same approximations that lead to Lindblad-Gorini-Kossakowski-Sudarshan (LGKS) master equations~\cite{lindblad1976a,gorini1976a} for an undriven system also for its driven version~\cite{Kohler_1997,GRIFONI1998229,gelbwaser_klimovsky2013a,szczygielski2013a,Szczygielski_2014}.
In the interaction picture, the application of Born-, Markov-, and secular approximations generically leads to LGKS master equations that come in handy due to their stability and -- despite the lack of thermalization~\cite{Shirai_2015,Shirai_2016} -- their transparent thermodynamic interpretation~\cite{bulnes_cuetara2015a}.
However, it has also frequently been noted that in particular the involved secular approximation may lead to strong artifacts~\cite{restrepo2019a,luo2020a}.
An intuitive explanation for this is that the level spacing in the extended (Floquet) space may become very small, which conflicts with the secular approximation~\cite{PhysRevE.79.051129}.
Therefore, traditional Floquet-LGKS approaches have been questioned in many works~\cite{restrepo2019a,schnell2020a}.

Already for undriven systems, it can be argued that the (non-LGKS) Redfield equation  applied in the proper regime only leads to small violations of density matrix properties 
in trade for a closer description of physical reality~\cite{hartmann2020a}.
As a practical benefit, improper density matrices can then be used as an indicator for leaving the region of validity of such perturbative schemes, whereas such a witness is missing for LGKS-based approaches.

\begin{figure}[bh]
\includegraphics[width=0.45\textwidth]{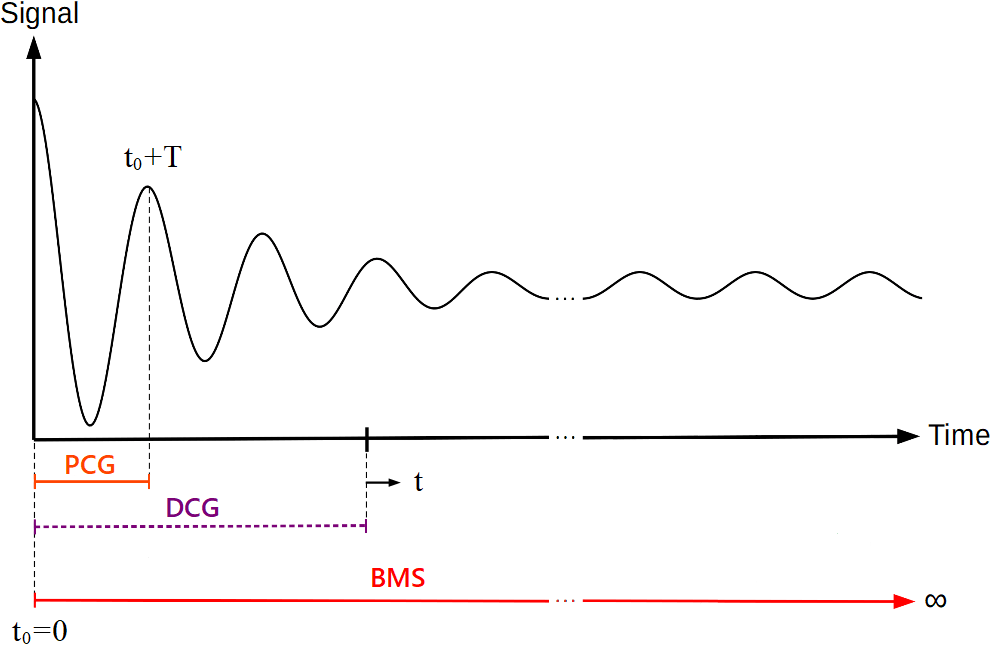}
\caption{(Color Online) Sketch illustrating the different coarse-graining timescales to derive a master equation in the weak coupling limit. For period coarse graining (PCG) (orange) the coarse-graining extends over one period of the driving field $\tau=T$. For dynamical coarse-graining (DCG) (purple) the coarse-graining time is chosen dynamically as the physical time. For $\tau\to\infty$ all methods recover the Born-Markov-secular (BMS) limit (red).}
\label{FIG:sketch_comparison_of_methods}
\end{figure}

Another viewpoint is to preserve the formal LGKS form of the generator while improving in some ways the involved approximations~\cite{whitney2008a}.
There are various routes to achieve this, e.g. formally restoring the LGKS form by discarding its subspace with negative eigenvalues~\cite{kirsanskas2018a,ptaszynski2019a}, by performing only a partial secular approximation~\cite{farina2019a,cattaneo2020a}, or by taking the effects of the reservoir on the system pointer basis into account~\cite{schultz2009a,schultz2010a,teretenkov2021a,trushechkin2021a}.
While these approaches apply to undriven systems, we would like to focus on an approach that can be easily combined with periodic driving.

Such an approach that can also be  well-motivated using a microscopic derivation is temporal coarse-graining~\cite{Vacchini_2000,Lidar_2001,Schaller_2008,schaller2009a,benatti2009a,benatti2010a,osti_388376,schaller2009a,majenz2013a,rivas2017a,rivas2019a,hartmann2020a,mozgunov2020a}.
Intuitively, it can be understood as an approximate Markovian representation of the reduced exact solution over a chosen coarse-graining time interval.
The freedom in choosing the coarse-graining interval is exemplified in Fig.~\ref{FIG:sketch_comparison_of_methods} and shall be the subject of this paper with a focus on periodically driven systems.
We begin with a brief introduction into the method in Sec.~\ref{SEC:method} and then compare various dissipators for three examples of driven open systems in Secs.~\ref{SEC:pure_dephasing},~\ref{SEC:circular_driving}, and~\ref{SEC:fast_driving}.


\section{Coarse-graining methods for periodically driven open systems}\label{SEC:method}

The starting point of our considerations is a representation of the Hamiltonian of the full universe in the form
\begin{align}
    H(t) = H_S(t) + \sum_\alpha A_\alpha \otimes B_\alpha + H_B
\end{align}
with periodically driven system Hamiltonian $H_S(t+T)=H_S(t)$, system and bath coupling operators $A_\alpha$ and $B_\alpha$, respectively, and reservoir Hamiltonian $H_B$.
We do not necessarily require the coupling operators to be individually hermitian, {but the total interaction Hamiltonian is hermitian $H_I = \sum_\alpha A_\alpha \otimes B_\alpha = H_I^\dagger$.
Furthermore, we assume the system coupling operators $A_\alpha$ to be normalized with respect to some norm, such that unit and strength of the interaction is carried by the bath coupling operators $B_\alpha$.
}
Since system and bath operators act on different Hilbert spaces, we assume from the beginning that $[H_S,H_B]=0$ and also $[A_\alpha,B_\beta]=0$, $[A_\alpha, H_B]=0$, $[B_\alpha, H_S]=0$.
In an interaction picture with respect to $H_0 = H_S(t) + H_B$ (marked by bold symbols), we can approximate the time evolution operator $\f{U}(t_0+\tau,t_0)$.
The coarse-graining approach originates from the attempt to match the reduced evolution of the open quantum system with a time-local master equation
\begin{align}\label{EQ:cgdef}
    e^{\f{{\cal L}_\tau} \tau} \rho_S^0 \stackrel{!}{=} \traceB{\f{U}(t_0+\tau,t_0) \rho_S^0 \otimes \rho_B^0 \f{U^\dagger}(t_0+\tau,t_0)}\,.
\end{align}
Here, we have already put an index ${\cal L}_\tau$ to the dissipation superoperator to highlight the fact that with a constant superoperator ${\cal L}$ it will in general not be possible to capture the exact reduced dynamics for all times, which would require a Kraus representation~\cite{kraus1971a}.
For weak system-reservoir interaction (or small propagation times $\tau$) we can use that $\f{U}(t_0+\tau,t_0)$ is close to the identity and $\f{{\cal L}}_\tau$ is small.
One can then expand the expressions on both sides and assuming furthermore that $\trace{B_\alpha \rho_B^0}=0$, one arrives at a defining equation for the coarse-graining generator, see App.~\ref{app:details} for details.
Choosing for simplicity $t_0=0$ it reads
\begin{align}\label{cg mastereq}
  \f{{\cal L}_\tau}\f{\rho_S} & =  -\ii \Big[\frac{1}{2\ii\tau} \int\limits^{\tau}_{0} dt_1 dt_2\sum_{\alpha \bar{\alpha}}C_{\alpha \bar{\alpha}}(t_1,t_2)\\
  &\qquad\qquad
  \times {\rm sgn}(t_1-t_2)\boldsymbol{A}_\alpha(t_1)\boldsymbol{A}_{\bar{\alpha}}(t_2),\boldsymbol{\rho_S}\Big]\nn
  &\qquad+\frac{1}{\tau}\int\limits^{\tau}_{0}dt_1 dt_2\sum_{\alpha \bar{\alpha}}C_{\alpha \bar{\alpha}}(t_1,t_2)\nn
  & \qquad
\times \Big[\boldsymbol{A}_{\bar{\alpha}}(t_2)\boldsymbol{\rho_S}\boldsymbol{A}_\alpha(t_1) -\frac{1}{2}\{\boldsymbol{A}_\alpha(t_1)\boldsymbol{A}_{\bar{\alpha}}(t_2),\boldsymbol{\rho_S}\}\Big]\,,\nonumber
\end{align}
with bath correlation functions $C_{\alpha \bar{\alpha}}(t_1,t_2)={\rm Tr_B} \{\boldsymbol{B_\alpha}(t_1)\boldsymbol{B_{\bar{\alpha}}}(t_2)\rho_B^0\}
= \trace{\f{B_\alpha(t_1-t_2)} B_\beta \rho_B^0}$, where the last equality holds when $[H_B,\rho_B^0]=0$.
{Here, the interaction picture representation of the system coupling operators is formally given by
$\f{A_\alpha}(t) = U_S^\dagger(t) A_\alpha U_S(t)$ with $U_S(t)=\mathcal{T}\left\{e^{-\ii \int_0^t H_S(t') dt'}\right\}$ and $\mathcal{T}$ denoting the time ordering operator.}
One can show that for any kind of driving and also for any fixed choice of $t_0$ and $\tau$, the above coarse-graining dissipator $\f{{\cal L}_\tau}$ is of LGKS form~\cite{mozgunov2020a,Schaller_2020}.
This contrasts the above generator from other coarse-graining approaches that arise from a temporal averaging of Redfield equations, which require subsequent secular-type approximations to reach an LGKS form~\cite{elouard2020a}.

{Thus, we can generally split the generator into a unitary part and a dissipative one}
\begin{align}
    \boldsymbol{\dot{\rho}_S} &=-\ii  \left[\f{H_{\rm LS}^\tau},\boldsymbol{\rho_S}\right]+ \f{\mathcal{D}^\tau} \f{\rho_S}\,,
\label{cg floquet mastereq}
\end{align}
which specifically for periodic driving have the single-integral expressions
\begin{align}
\label{EQ:CGx}
    \f{H_{\rm LS}^\tau}&= \int d\omega \sum_{\alpha \bar{\alpha}} \sigma_{\alpha\bar{\alpha}}(\omega) \frac{1}{2 \ii}\sum_{nn'}\sum_{abcd}\nn
    & \qquad \times f_{0}^\tau(\bar{E}_a-\bar{E_b}+n \Omega,\bar{E}_c-\bar{E}_d+n'\Omega,\omega)\nn
    &\qquad \times A_{\alpha,ab}^{n} A_{\bar{\alpha},dc}^{-n'}L_{ab} L_{dc}\,,\nn
\f{\mathcal{D}^\tau} \boldsymbol{\rho_S} &= \int d\omega \sum_{\alpha\bar{\alpha}} \gamma_{\alpha\bar{\alpha}} (\omega)\sum_{nn'}\sum_{abcd}\nn
&\qquad\times f_{0}^\tau(\bar{E}_a-\bar{E}_b+n \Omega,\bar{E}_c-\bar{E}_d+n' \Omega,\omega)\nn
&\qquad \times A_{\alpha,ab}^{+n} A_{\bar\alpha,dc}^{-n'}\nn
    & \qquad \times
    {\left[L_{dc} \boldsymbol{\rho_S} L_{ab} -\frac{1}{2}\left\{L_{ab} L_{dc} ,\boldsymbol{\rho_S}\right\}\right]
    }\,.
\end{align}
{Here, the first term converges for large $\tau$ to the LGKS Lambshift term, and the second term denotes the dissipative influence of the reservoir.
The functions $\sigma_{\alpha\bar\alpha}(\omega)$ as well as $\gamma_{\alpha\bar\alpha}(\omega)$ denote odd and even Fourier transforms of the reservoir correlation functions~\eqref{EQ:corrfuncdef}, $f_{0}^\tau(E_a,E_b,\omega)$ is a nascent $\delta$-function, the $A_{\alpha,ab}^n$ are Fourier components of the system coupling operator matrix elements, the 
quasienergies $\bar{E}_a \in [-\Omega/2,+\Omega/2)$ are chosen in the first Brillouin zone defined by the period of the driving $\Omega=2\pi/T$,
and the Lindblad operators $L_{ab}$ generate transitions between Floquet Hamiltonian eigenstates, see App~\ref{app:details} for the precise definitions.}

The different coarse-graining schemes sketched in Fig.~\ref{FIG:sketch_comparison_of_methods} may lead to different LGKS-type dissipators.
{
For later reference, we coin the coarse-graining schemes {\em dynamical coarse-graining}~\cite{Schaller_2008} (DCG) with $t_0=0$ and $\tau=t$ and
{\em period-coarse-graining} (PCG) with $t_0=0$ and $\tau=T$.
Details for these schemes are found in Appendices~\ref{APP:dcg} and~\ref{APP:cg1},  respectively.
Furthermore}, we remark that the limit of very large coarse-graining times $\tau\to\infty$ allows for significant simplifications, see App.~\ref{APP:longterm}, which however do not always coincide with the frequently-used Born-Markov-secular {(BMS)} Floquet master equation exposed in App.~\ref{APP:bms} and a further simplified variant thereof -- {the Born-Markov-ultrasecular (BMU)} master equation -- in App.~\ref{APP:bmu}.

{
Our problem is defined by three physical timescales.
We will be concerned with driven two-level systems, where the natural internal time scale of the system dynamics is defined by their level splitting $\Delta E$ via the relation $\tau_{\rm int}=2\pi/\Delta E$.
Then, we have the obvious timescale given by the period of the external driving $\tau_{\rm drv}=2\pi/\Omega=T$.
Finally, the reservoir timescales are reflected in the decay timescale of the reservoir correlation function $\tau_{\rm dec}=2\pi/\omega_{\rm c}$, where $\omega_{\rm c}$ is a cutoff in the reservoir spectral density.
We are going to investigate how the technical timescale -- the coarse-graining time $\tau$ -- can be optimally chosen in dependence of the physical ones.
}

In what follows, we will compare the solutions to various periodically driven problems that are based on these different coarse-graining approaches, i.e., for the DCG approach we evaluate Eq.~(\ref{EQ:CGx}) with $\tau=t$, for the {PCG approach} we use Eq.~(\ref{EQ:CGx}) with $\tau=T$, whereas for the BMS and BMU results we use Eqns.~(\ref{EQ:bms}) and~(\ref{EQ:bmu}), respectively.
Furthermore, in our calculations we will for simplicity neglect the Hermitian correction term to the Hamiltonian $\f{H_{\rm LS}^{\ldots}}$ throughout.
Whereas for the particular pure-dephasing model considered below this term has no effect anyways, in general it can only be neglected in the weak-coupling regime where $\abs{H_{\rm LS}^{\ldots}} \ll \abs{H_S}$.


\section{Pure-dephasing models}\label{SEC:pure_dephasing}

By the term pure-dephasing models we summarize models where the interaction commutes with the system Hamiltonian at all times, i.e., system and reservoir cannot exchange energy.
The Hamiltonian of a general driven-system pure-dephasing model with a reservoir of bosonic oscillators is then given by:
\begin{align}\label{EQ:pure_dephasing}
    H(t) &= H_S(t) 
    + A \otimes \sum_k \left(h_k b_k + h_k^* b_k^\dagger\right) 
    + \sum_k \omega_k b_k^\dagger b_k\,,\nn
&\left[H_S(t), A\right] =0\,.
\end{align}
The assumption of commuting coupling operator $A$ and system Hamiltonian $H_S(t)$ allows to solve the dynamics exactly, {see App.~\ref{APP:pd}}.

\subsection{Exact solution}

Specifically for a driven two level system with \mbox{$H_S = \sigma^z \left[\frac{\Delta}{2}+ \lambda \cos(\Omega t)\right]$} {with the single system coupling operator} $A=\sigma^z$ we obtain that the populations in the $\sigma^z$ eigenbasis remain constant
\begin{align}\label{EQ:pdp_pop}
    \rho_{S,00}(t)=\rho_{S,00}^0\,,\qquad \rho_{S,11}(t)=\rho_{S,11}^0\,,
\end{align}
whereas the coherences {decay oscillatorily} according to
\begin{align}\label{EQ:pdp_exact}
       \rho_{S,10}(t)&= e^{+\ii\left(\Delta \cdot t+2 \frac{\lambda}{\Omega}\sin(\Omega t)  \right)}\\
       & \qquad \times \exp\left\{ -\frac{4}{\pi}\int\limits_{-\infty}^\infty d\omega \gamma(\omega)\frac{\sin^2 \left( \frac{\omega t}{2}\right)}{\omega^2} \right\}\rho_{S,10}^0\,.\nonumber
\end{align}
Here, {the Fourier transform of the single reservoir correlation function is} $\gamma(\omega) = \Gamma(\omega)[1+n_B(\omega)]$ with spectral density $\Gamma(\omega)\equiv 2\pi \sum_k \abs{h_k}^2 \delta(\omega-\omega_k)$ (analytically continued via $\Gamma(-\omega)=-\Gamma(\omega)$ to the complete real axis) and the Bose distribution is $n_B(\omega) = [e^{\beta\omega}-1]^{-1}$.
In absence of driving ($\lambda\to 0$), this falls back to the known pure-dephasing solution of the spin-boson model, see e.g. Ref.~\cite{Lidar_2001}.
{Additionally, we remark that one may also obtain the above result from the undriven model by a gauge transform.}

\subsection{Master equation solutions}

We can now compare this exact solution with the various approximate approaches discussed before by {dropping the indices as above, i.e., using $A_1=A=\sigma^z$
and $\gamma_{11}(\omega)=\gamma(\omega) = \Gamma(\omega)[1+n_B(\omega)]$ for coupling operator and Fourier transform of the reservoir correlation function.
Notably, the matrix elements of the coupling operators become rather trivial
$A_{1,ab}^{n} = \delta_{n,0}\delta_{ab}\bra{\bar{a}} \sigma^z \ket{\bar{a}}$,
where $\ket{\bar{a}}$ are Floquet Hamiltonian eigenstates.}
To compute them, it suffices to realize that for this simple problem the system Floquet Hamiltonian is just $\bar{H}=\frac{\Delta}{2} \sigma^z$, such that its eigenstates are trivial. 
All master equations capture the constant populations~\eqref{EQ:pdp_pop}.

The coarse-graining solution for fixed coarse-graining time $\tau$ predicts for the coherences a decay according to:
\begin{align}\label{EQ:pd_cg}
    \rho_{cg,10}(t)&=e^{+\ii\left(\Delta \cdot t+2 \frac{\lambda}{\Omega}\sin (\Omega t)\right)}\\
  & \qquad \times \exp\left\{-\frac{4}{\pi}{\frac{t}{\tau}}\int\limits_{-\infty}^{\infty} \gamma(\omega) \frac{\sin^2(\frac{\omega{ \tau}}{2})}{\omega^2}d\omega\right\}\rho^0_{10}\,,\nonumber
\end{align}
where the PCG approach is recovered for $\tau=2\pi/\Omega$.

Clearly, for DCG ($\tau=t$), the above equation exactly reproduces the exact solution~\eqref{EQ:pdp_exact}.

Furthermore, for this model the BMS and BMU solutions coincide:
\begin{align}
  \rho_{\rm BM,10}(t)=e^{+\ii\left(\Delta \cdot t+2 \frac{\lambda}{\Omega}\sin (\Omega t)\right)}e^{-2 \gamma(0) t}\rho^0_{10}\,,   
    \label{pd_bms}
\end{align}
which can also be obtained by virtue of~\eqref{EQ:deltafunc} from Eq.~\eqref{EQ:pd_cg} in the limit $\tau\to\infty$.

\subsection{Comparison}

\begin{figure}
\centering
\includegraphics[width=0.45\textwidth,clip=true]{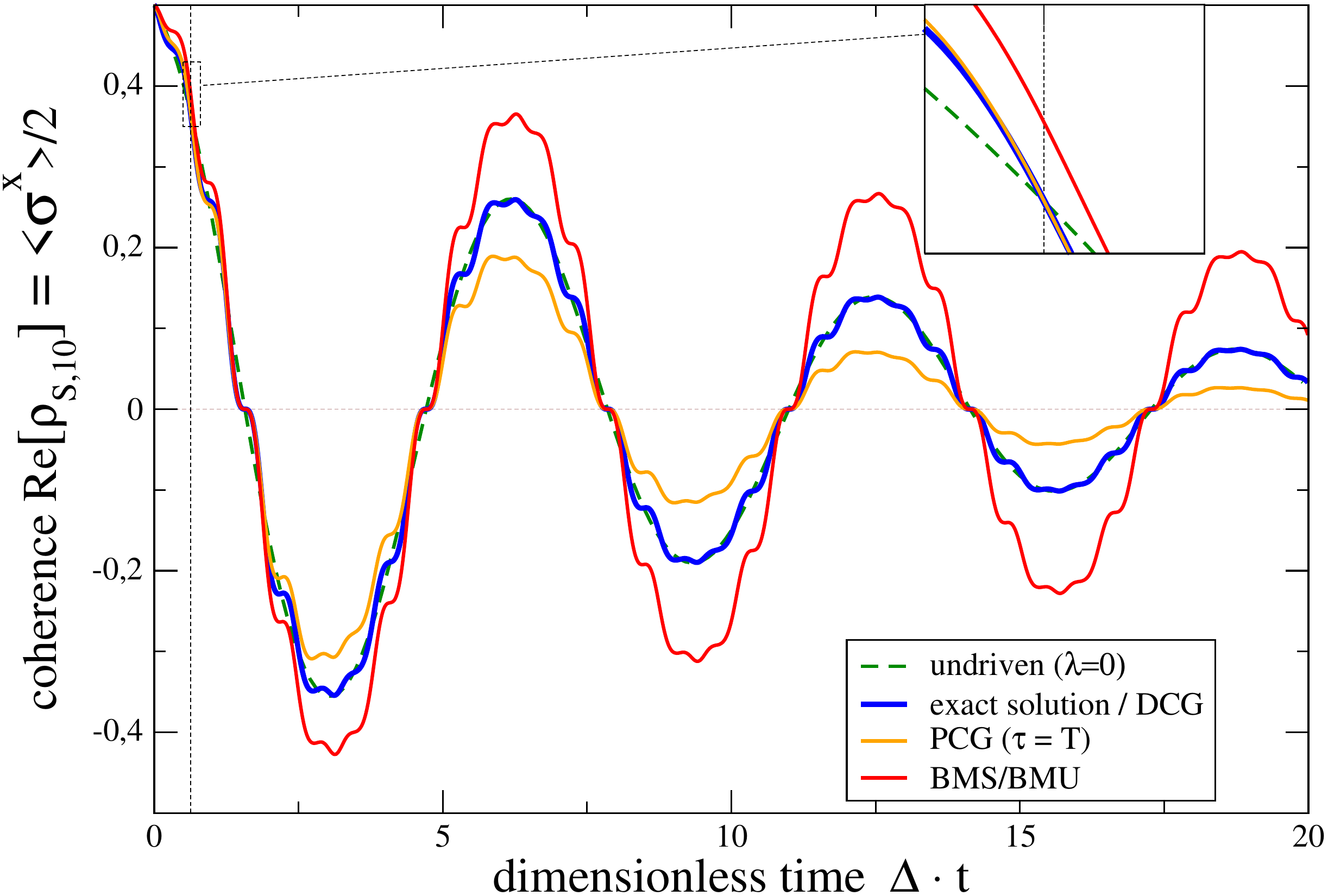}
\caption{(Color Online) Plot  of  $\expval{\sigma^x}/2$ comparing coarse-graining PCG for $\tau=T=\frac{2 \pi}{\Omega}$ (orange) and the BMS/BMU solution (red) to the exact analytical solution (blue) {and the exact solution in absence of driving (dashed green)}. BMS and BMU solutions are identical, and the DCG approach ($\tau=t$) reproduces the exact solution. 
Inset: By construction, PCG (orange) and DCG solutions (blue) intersect at
$t=T$ (vertical dashed line). Parameters: $\Omega=10\;\Delta $, $\lambda=\frac{1}{2}\; \Delta$, $\Gamma(\omega) = \Gamma_0 \omega e^{-\abs{\omega}/\omega_{\rm c}}$ with $\Gamma_0=0.05$, $\omega_c=20 \;\Delta$, $\beta \, \Delta=1$, $\rho_{\rm S}^0 = 1/2(\f{1}+\sigma^x)$.
\label{FIG:pure_dephasing}}
\end{figure}

We analytically find that the DCG approach (adaptively choosing $\tau=t$ and $t_0=0$ in Eq.~(\ref{EQ:CGx})) always reproduces the exact solution.
We also find analytically from the trivial structure of the model that BMS~(\ref{EQ:bms}) and BMU~(\ref{EQ:bmu}) solutions must concide.
In Fig.~\ref{FIG:pure_dephasing} one can see that the additional driving modifies the decay of coherences by super-imposing small additional oscillations to {the undriven solution (green dashed)}.
Apart from that, we see that
the exact reduced dynamics cannot be reproduced by a single Markovian generator (like PCG, BMS or BMU).
{Fig.~\ref{FIG:pure_dephasing} has been computed for parameters with $\omega_c > \Omega \gg \Delta$ ($\tau_{\rm dec} < \tau_{\rm drv} \ll \tau_{\rm int})$.

When we consider slower drivings $\omega_c \gg \Omega = \Delta$ ($\tau_{\rm dec} \ll  \tau_{\rm drv} = \tau_{\rm int}$), we observe that also the PCG method converges towards the exact solution, see Fig.~\ref{FIG:pure_dephasing1}, whereas the BMS approach fails to capture the exact dynamics.
\begin{figure}
\centering
\includegraphics[width=0.45\textwidth,clip=true]{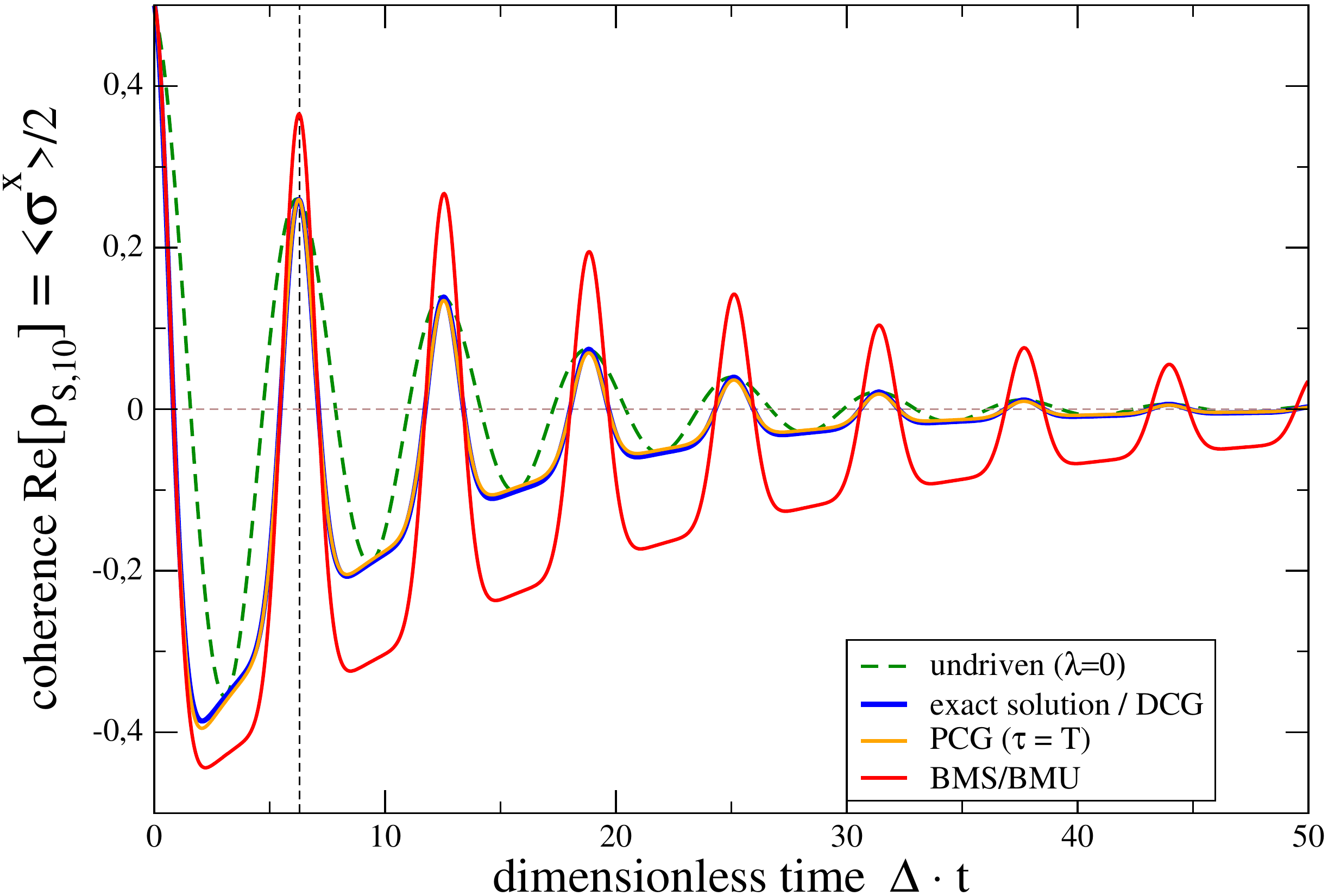}
\caption{(Color Online) Analogous to Fig.~\ref{FIG:pure_dephasing}, but for slower driving $\Omega=\Delta$. PCG and exact solution agree much better.
\label{FIG:pure_dephasing1}}
\end{figure}
The failure of the BMS approach in this limit is not too surprising, given that it is typically derived for fast driving. 
The fact that for slow driving it suffices to coarse-grain over the first period (PCG) to get a sufficiently accurate dissipator is a consequence of the relatively high temperatures chosen.
This can be formally seen by performing a variable transform $\omega=\omega' t/\tau$ in~\eqref{EQ:pd_cg} and using that in the considered range of parameters $\gamma(\Omega t/(2\pi) \omega) \approx \Gamma_0/\beta \approx \gamma(\omega)$.
For lower temperatures (not shown) and other models (see below) we do not see such a clear tendency.
}

Since pure-dephasing-type models are a very specific model class, we consider models that allow the exchange of energy between system and reservoir below.


\section{Circular driving}\label{SEC:circular_driving}

As a second model, we consider a periodically driven open two level system of the form
\begin{align}\label{EQ:circ_driv}
    H(t)&=\frac{\Delta }{2}\sigma^z+P\sigma^+ e^{-\ii \Omega t}+P^*\sigma^-e^{+\ii\Omega t}\nn
    &\quad +\sum_k(\sigma^+ h_k b_k+\sigma^-h_k^*b_k^\dagger)+\sum_k \omega_k b_k^\dagger b_k\,,
\end{align}
where $P$ denotes a driving amplitude.
{For $P=0$ and a vacuum reservoir, the model is exactly solvable~\cite{garraway1997a,book_breuer}, but we are here interested in the case of finite driving amplitudes but weak system-reservoir coupling.
We therefore interpret the DCG solution as benchmark for this model, see App.~\ref{APP:convergence}.

Before treating the model with our standard approaches, we demonstrate that the model can be mapped to a different frame where the Hamiltonian becomes time-independent.
In this time-independent frame, a standard master equation can be derived that is equivalent to the Floquet (BMS) master equation in the original frame.
}

\subsection{{Mapping to a time-independent frame}}

By employing the unitary gauge transform 
\begin{align}\label{EQ:gauge}
    V(t)= \exp\left[-\ii \left( \frac{\Omega t}{2}\sigma^z+\Omega t \sum_k b_k^\dagger b_k \right) \right]
\end{align}
one can see from the observations
\begin{align}
    V^\dagger(t) \sigma^z V(t) &= \sigma^z\,,\qquad
    V^\dagger(t) \sigma^\pm V(t) = \sigma^\pm e^{\pm\ii\Omega t}\,,\nn
    V^\dagger(t) b_k V(t) &= e^{-\ii\Omega t} b_k\,,\qquad
    V^\dagger(t) b_k^\dagger V(t) = e^{+\ii\Omega t} b_k^\dagger\,,
\end{align}
that the total Hamiltonian becomes time-independent under the transformation
\begin{align}\label{EQ:gauge_transform}
    V^\dagger(t) H(t) V(t) &= \frac{\Delta }{2}\sigma^z+P\sigma^++P^*\sigma^-\\
    &\quad +\sum_k(\sigma^+ h_k b_k+\sigma^-h_k^*b_k^\dagger)+\sum_k \omega_k b_k^\dagger b_k\,.\nonumber
\end{align}
Accordingly, by transforming into the corresponding picture, the expectation value of an observable $O$ can be written as
\begin{align}
    \expval{O} = \trace{O U(t) \rho_0 U^\dagger(t)} 
    \equiv \trace{\tilde{O}(t) \tilde{U}(t) \rho_0 \tilde{U}^\dagger(t)}
\end{align}
with the effective time evolution operator $\tilde{U}(t) = V^\dagger(t) U(t)$ and transformed observable
$\tilde{O}(t) = V^\dagger(t) O V(t)$.
The effective time evolution operator evolves according to a time-independent picture
\begin{align}
    \frac{d}{dt} \tilde{U}(t) &= -\ii \tilde{H} \tilde{U}(t)\,,\\
    \tilde{H} &= V^\dagger(t) H(t) V(t)+\ii \dot{V} V\nn
    &= \frac{\Delta-\Omega}{2} \sigma^z + P \sigma^+ + P^* \sigma^-\nn
    &\quad+\sum_k(\sigma^+ h_k b_k+\sigma^-h_k^*b_k^\dagger)+\sum_k (\omega_k-\Omega) b_k^\dagger b_k\,,\nonumber
\end{align}
such that we can write $\tilde{U}(t) = e^{-\ii \tilde{H} t}$.

Therefore, under the gauge transform~\eqref{EQ:gauge_transform}, the time evolution is just that of a time-independent open two-level system.
In this picture, we can derive the BMS master equation for an undriven system.
Formally, this master equation looks like Eq.~(\ref{EQ:bms}) with keeping only the $n=0$ term and instead of the Floquet Hamiltonian eigenstates and eigenvalues using the eigenstates and eigenvalues of 
$\tilde{H}_S = \frac{\Delta-\Omega}{2} \sigma^z + P \sigma^+ + P^* \sigma^-$.

{In the last line of the above equation, we can see that the reservoir single-particle energies are shifted, which eventually also affects the Kubo-Martin-Schwinger~\cite{kubo1957a,martin1959a,kossakowski1977a} relation obeyed by the reservoir correlation function.
Therefore, the resulting master equation} does not thermalize even in the time-independent picture but reaches some nonequilibrium steady state, which -- after transforming back to the original picture -- exhibits a time-dependence with the period of the driving.
{This becomes more evident in a simpler long-term analysis:}
For non-degenerate eigenvalues of the {effective system Hamiltonian $\tilde{H}_S$}, the populations decouple from the coherences and the evolution is given by a simple rate equation:
$\dot{\tilde{\rho}}_{S,aa}=\sum_{b}\gamma_{ab,ab}\tilde{\rho}_{S,bb}-\sum_b \gamma_{ba,ba}\tilde{\rho}_{S,aa}$
with transition rates from energy level $b$ to $a$ given by
$\gamma_{ab,ab}$.
The steady state solution is then given by
\begin{align}\label{EQ:ss_rate}
    \bar{\tilde{\rho}}_S&=P_-\ket{-}\bra{-}+(1-P_-)\ket{+}\bra{+}\,,\nn
    P_-&=\frac{\gamma_{-+,-+}}{\gamma_{-+,-+}+\gamma_{+-,+-}}\,,
\end{align}
where $\ket{-}$ and $\ket{+}$ denote the ground and excited state of {$\tilde{H}_S$}, respectively.
For our example the two transition rates read
\begin{align}
   \gamma_{-+,-+}&=\tilde{\gamma}_{12}(E_+ -E_-)|\bra{-}\sigma^-\ket{+}|^2\nn
   &\qquad+\tilde{\gamma}_{21}(E_+ -E_-)|\bra{-}\sigma^+\ket{+}|^2\,,\nn
   \gamma_{+-,+-}&=\tilde{\gamma}_{12}(E_- -E_+)|\bra{+}\sigma^-\ket{-}|^2\nn
   &\qquad+\tilde{\gamma}_{21}(E_- -E_+)|\bra{+}\sigma^+\ket{-}|^2\,,
\end{align}
with $\tilde{\gamma}_{12}(\omega) = \Theta(\omega+\Omega) \Gamma(\omega+\Omega) [1+n_B(\omega+\Omega)]$
and $\tilde{\gamma}_{21}(\omega) = \Theta(-\omega+\Omega) \Gamma(-\omega+\Omega) n_B(-\omega+\Omega)$ exemplifying the broken Kubo-Martin-Schwinger relation $\frac{\tilde{\gamma}_{21}(-\omega)}{\tilde{\gamma}_{12}(+\omega)}=e^{-\beta(\omega+\Omega)}$.
In this picture, system observables can be computed via 
\begin{align}\label{EQ:asymptotic_state}
    \expval{O} \to \ptrace{S}{e^{+\ii\Omega t/2 \sigma^z} O e^{-\ii\Omega t/2 \sigma^z} \tilde{\rho}_S(t)}\,,
\end{align}
where for large times we can insert $\tilde{\rho}(t) \to \bar{\tilde{\rho}}_S$ as given by Eq.~(\ref{EQ:ss_rate}).

{We find numerically that the solutions derived in this way fully agree with the BMS solutions in the original frame (solid red curves in Figs.~\ref{FIG:circular_SX} and~\ref{FIG:circular_SZ}).
This is not surprising as the gauge transformation that we used should not change the general outcome.
In addition, we checked that the simplified long-term dynamics agrees with the full time-dependent solution for large times.
}

\subsection{Master equation solutions}

We now go back to the original Schr\"odinger picture representation of Eq.~(\ref{EQ:circ_driv}), where we can identify system coupling operators $A_1 = \sigma^+$ and $A_2 = \sigma^-$.
Additionally, the non-vanishing correlation functions become $\gamma_{12}(\omega) = \Theta(\omega) \Gamma(\omega) [1+n_B(\omega)]$ and $\gamma_{21}(\omega) = \Theta(-\omega) \Gamma(-\omega) n_B(-\omega)$.
{With a gauge transform analogous to~\eqref{EQ:gauge} but only involving the system parts one can find the exact time evolution operator of the driven system alone, and by demanding periodicity of the kick operator one finds that} kick-operator and Floquet Hamiltonian are given by
\begin{align}
    \bar{H} &= \frac{\Delta-\Omega}{2} \sigma^z + P \sigma^+ + P^* \sigma^- - \frac{\Omega}{2}\f{1}\,,\nn
    U_{\rm kick}(t) &= \exp\left(-\ii\frac{\Omega t}{2} (\f{1}+\sigma^z)\right)\,.
\end{align}
This allows us to compute the coarse-graining dissipators DCG (App.~\ref{APP:dcg}) and PCG (App.~\ref{APP:cg1}) as well as BMS (App.~\ref{APP:bms}) and BMU (App.~\ref{APP:bmu}) dissipators.

\subsection{Comparison}

\begin{figure}
\centering
\includegraphics[width=0.45\textwidth,clip=true]{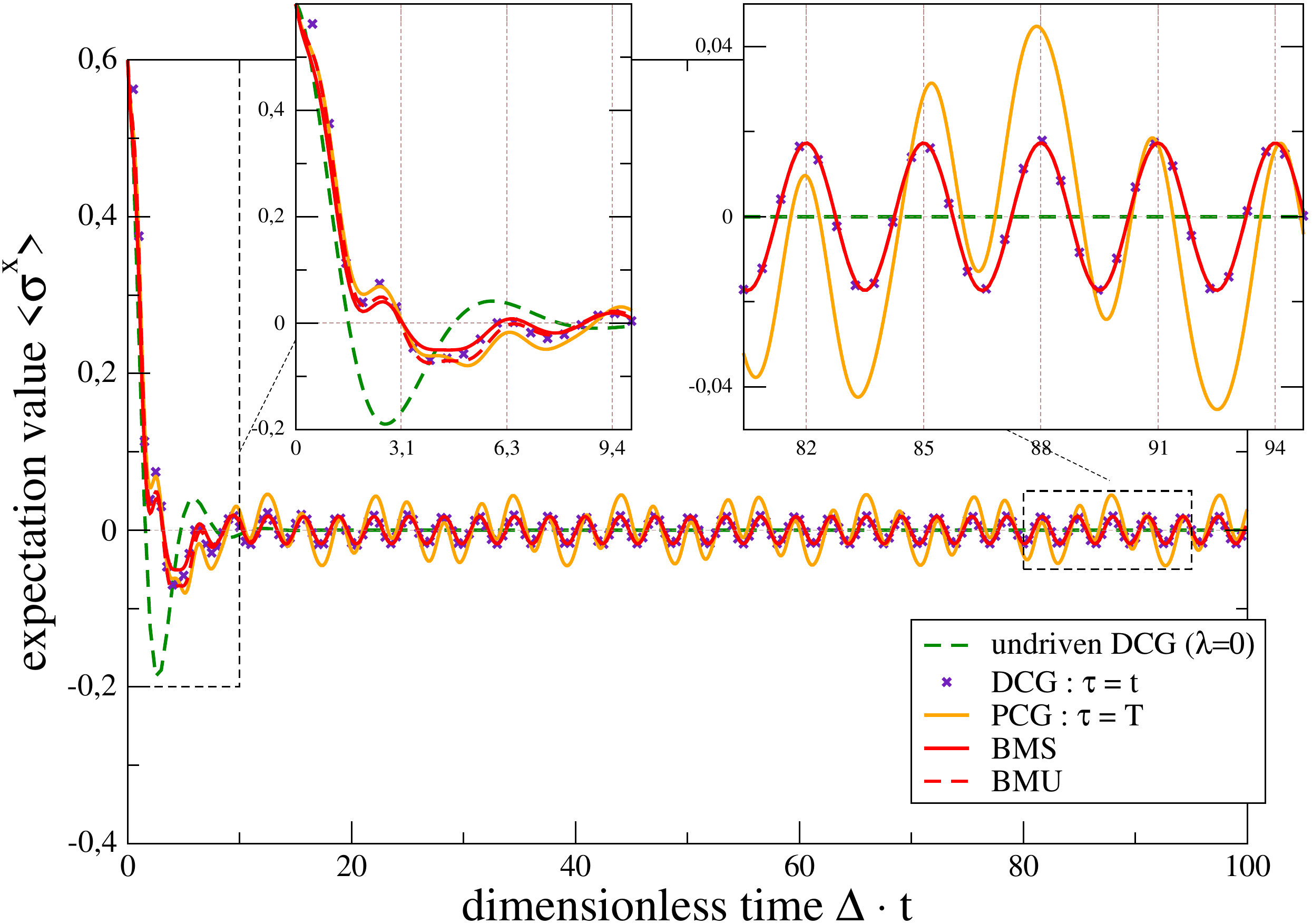}
\caption{(Color Online) Plot  of  the  time-dependent  expectation value  $\expval{\sigma^x}$ according to the {undriven DCG solution (dashed green)}, driven DCG (purple crosses), PCG (orange), and BMS (solid red) and BMU (dashed red) solutions. The gridlines mark multiples of the driving period $T$. At long times, DCG, BMS, and BMU solutions all agree (right inset). 
For short times, the BMS and BMU solutions differ from each other and also from the DCG (benchmark) solution. Parameters: $\Omega=2 \; \Delta$, $P=\frac{1}{2}\; \Delta$ (with Floquet energies ($\bar{E}_\pm = \pm \frac{2-\sqrt{2}}{2}\Delta$) in the first Brillouin zone), $\beta \, \Delta=0.1$, $\Gamma_0=0.05$, $\omega_c=15 \; \Delta$,
$\rho_S^0 = 1/2[\f{1}+0.6 \sigma^x + 0.4 \sigma^z]$.}
\label{FIG:circular_SX}
\end{figure}

\begin{figure}
\centering
\includegraphics[width=0.45\textwidth,clip=true]{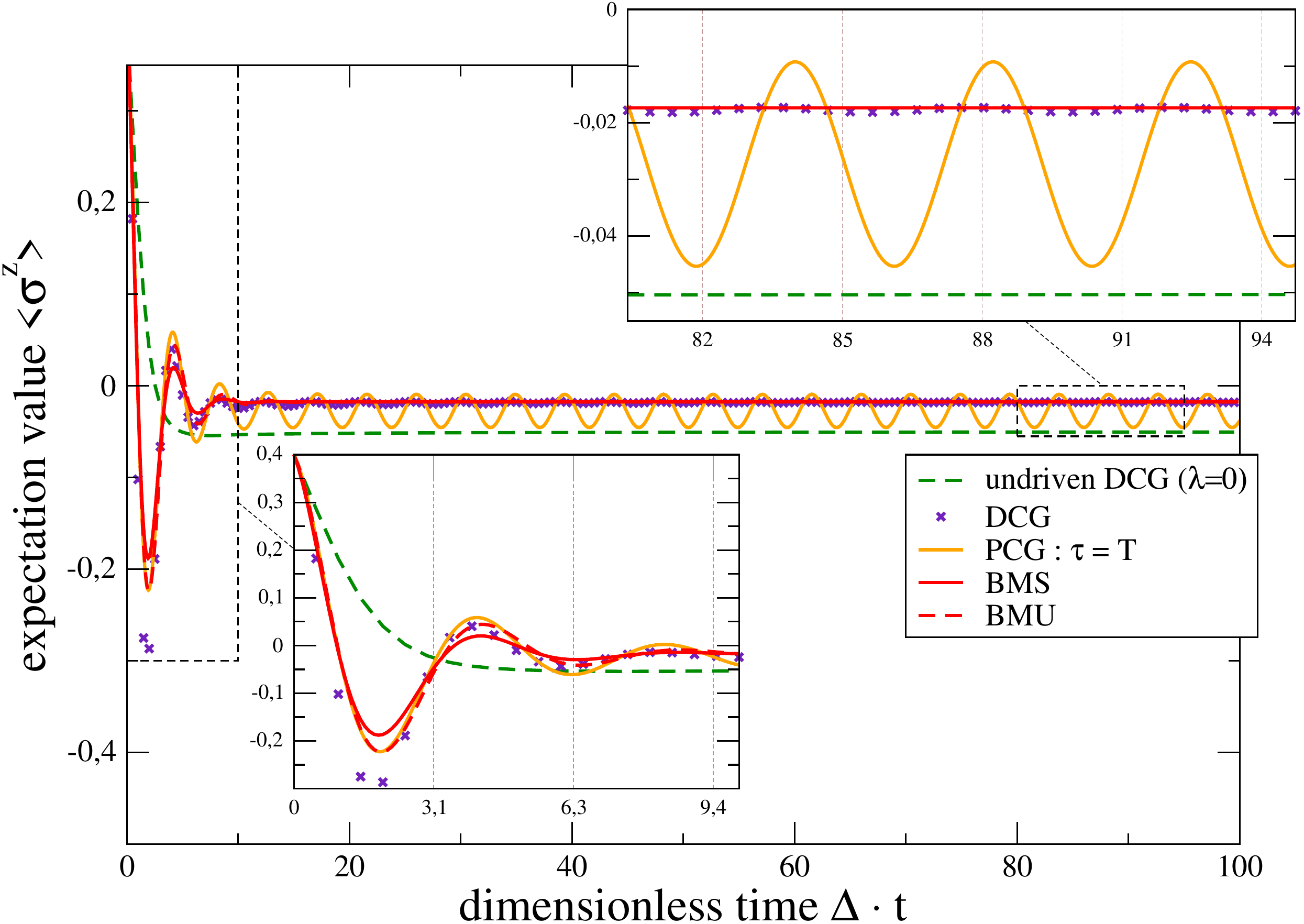}
\caption{(Color Online) Analogous to Fig.~\ref{FIG:circular_SX} but for $\expval{\sigma^z}$. By construction, for short times the DCG solution is superior to the other perturbative methods, but again DCG, BMS, and BMU solutions converge to for large times.
}
\label{FIG:circular_SZ}
\end{figure}

We compare the resulting dynamics with the previously discussed asymptotic solution in Fig.~\ref{FIG:circular_SX}.
{At steady state, we see a convincing agreement of the BMS and BMU (solid and dashed red) approaches with the DCG approach.
Additionally, we checked that the BMS approach fully agrees with the master equation solution from the time-independent picture (not shown).
In the short-term limit however, some differences become visible (left inset).
In contrast, the PCG solution (orange) captures the short-term evolution but fails to match the DCG solution for large times. 

Analogous results hold for other expectation values, see Fig.~\ref{FIG:circular_SZ}.
}
Here one finds strong differences only transiently, where for short times the benchmark DCG approach by construction must approximate the unknown exact solution for the chosen parameters.

{These figures have been computed for $\omega_c\gg\Omega>\Delta$, i.e., for $\tau_{\rm dec} \ll \tau_{\rm drv} < \tau_{\rm int}$, which explains the good long-term agreement of the BMS and BMU solutions.

We also explored the regime of faster driving ($\omega_c > \Omega \gg \Delta$), where the BMS solution performs also transiently better and the regime of slower driving ($\omega_c \gg \Omega\approx\Delta$), where the BMS solution performed (transiently) worse. 
However, unlike Sec.~\ref{SEC:pure_dephasing}, the PCG approach did not approximate the DCG results very well for all times and observables at slow drivings.
}


\section{Fast-Driving solution}\label{SEC:fast_driving}

As a last example, we consider systems where the driving breaks the pure-dephasing character of the system
\begin{align}\label{EQ:fast_driving}
 H(t)&= H_S^0+ \lambda \cos(\Omega t) C + A \otimes \sum_k(h_k b_k+ h_k^*b_k^{\dagger})\nonumber\\&\qquad +\sum_k\omega_k b_k^{\dagger} b_k\,,
\end{align}
with system Hamiltonian $H_S^0$ and system coupling operator $A=A^\dagger$ fulfilling only the constraint $[H_S^0, A]=0$ and system driving operator $C=C^\dagger$ with in general $[H_S^0, C]\neq 0$.
Under a naive rotating wave approximation (RWA), the contribution of the driving vanishes completely, and in case of a qubit the exact solution~\eqref{EQ:pdp_exact} from the previous section in absence of driving ($\lambda=0$) applies for all values of the system-reservoir coupling strength.
However, to go beyond the naive RWA, we transform into an interaction picture with respect to the driving
\begin{align}\label{EQ:transform_driving}
    U_1(t) = e^{-\ii \lambda/\Omega \sin(\Omega t) C}\,,
\end{align}
and in this picture (marked by a tilde e.g. via $\tilde{A}(t) = U_1^\dagger(t) A U_1(t)$) the transformed Hamiltonian assumes a time-dependent pure-dephasing form
\begin{align}\label{EQ:transformed_fastdriving}
    \tilde{H}(t) &= \tilde{H}_S^0(t) + \tilde{A}(t) \otimes \sum_k(h_k b_k+ h_k^*b_k^{\dagger})
+\sum_k\omega_k b_k^{\dagger} b_k\,,
\end{align}
where we still have the equal-time commutation $\left[\tilde{H}_S^0(t), \tilde{A}(t)\right]=U_1^\dagger(t) \left[H_S^0, A\right] U_1(t)=0$.
Nevertheless, we cannot apply the naive polaron treatment of App.~\ref{APP:pd}, since now also the coupling operator has picked up a periodic time-dependence.
{Then, the polaron transform would become time-dependent as well, leading to an additional term in the effective Hamiltonian, such that the system-reservoir interaction would just be moved elsewhere.}

However, depending on the problem one may have the situation that also the time averages of the operators
\begin{align}
    \bar{H}_S^0 \equiv \frac{1}{T} \int_0^T \tilde{H}_S^0(t) dt\,,\qquad
    \bar{A} \equiv \frac{1}{T} \int_0^T \tilde{A}(t) dt
\end{align}
commute with each other $[\bar{H}_S^0, \bar{A}]=0$, and in this case the Hamiltonian after an RWA in the interaction picture
\begin{align}
    \bar{H}_{\rm tot} = \bar{H}_S^0 + \bar{A} \otimes \sum_k(h_k b_k+ h_k^*b_k^{\dagger})+\sum_k\omega_k b_k^{\dagger} b_k
\end{align}
is of pure-dephasing type and can be solved with standard methods, see App.~\ref{app fd} for details.

The transformation~\eqref{EQ:transform_driving} is of course not equivalent to the interaction picture representation. 
However, under the RWA in this transformed frame, the time evolution operator of the system becomes
$U_S(t) \approx U_1(t) e^{-\ii \bar{H}_S^0 t}$, such that we can identify approximations to kick operator $U_{\rm kick}(t) \approx U_1(t)$ and system Floquet Hamiltonian $\bar{H} \approx \bar{H}_S^0$, respectively.

\subsection{Fast driving benchmark}

Specifically, for a two level system with 
\begin{align}
H_S^0 = \frac{\Delta}{2} \sigma^z\,,\quad
A=\sigma^z\,,\quad C= \sigma^x,
    \label{tls fd}
\end{align}
we obtain with the Fourier decomposition
\begin{align}\label{EQ:coupling_op}
    \tilde{A}(t) &= U_1^\dagger(t) \sigma^z U_1(t)\nn
    &= \sigma^z \sum_{n=-\infty}^\infty {\cal J}_{2n}\left(\frac{2\lambda}{\Omega}\right) e^{\ii 2n\Omega t}\nn
&\qquad    -\ii \sigma^y \sum_{n=-\infty}^\infty {\cal J}_{2n+1}\left(\frac{2\lambda}{\Omega}\right) e^{\ii (2n+1)\Omega t}
\end{align}
that this method -- under the RWA in the transformed frame -- yields the following dynamics
\begin{align}\label{EQ:fastdriving}
        \expval{\sigma^x}&=\big[\cos\left(\mu_1 \Delta \cdot t\right)\expval{\sigma^x}_0-\sin\left(\mu_1\Delta \cdot t\right)\expval{\sigma^y}_0\big] \Sigma(t)\,,\nn
       \expval{\sigma^y} &=-\sin\left( \tfrac{2 \lambda}{\Omega}\sin(\Omega t)\right)\expval{\sigma^z}_0+\cos \left( \tfrac{2 \lambda}{\Omega}\sin(\Omega t)\right)\nn
     & \qquad \times \big[ \sin\left(\mu_1 \Delta \cdot t\right)\expval{\sigma^x}_0+\cos\left(\mu_1\Delta \cdot t\right)\langle \sigma^y\rangle_0\big] \Sigma(t)\,,\nn
       \expval{\sigma^z} &=+\cos\left( \tfrac{2 \lambda}{\Omega}\sin(\Omega t)\right)\expval{\sigma^z}_0+\sin \left( \tfrac{2 \lambda}{\Omega}\sin(\Omega t)\right)\nn
       & \qquad \times \big[ \sin\left(\mu_1 \Delta \cdot t\right)\expval{\sigma^x}_0+\cos\left(\mu_1\Delta \cdot t\right)\expval{\sigma^y}_0\big] \Sigma(t)\,,
\end{align}
with $\Sigma(t)\equiv\exp\left\{-\frac{4}{\pi}\left(\mu_1\right)^2\int\limits_{-\infty}^{\infty}\gamma(\omega) \frac{\sin^2\left(\frac{\omega t}{2}\right)}{\omega^2}d\omega \right\}$ and $\mu_1\equiv \mathcal{J}_0\left(\tfrac{2\lambda}{\Omega}\right)$ being given by the Bessel function of the first kind~\cite{abramowitz1970}.
Details on this can be found in App.~\ref{app fd} by inserting the expressions for a two level system in Eq.~(\ref{exp value fd}).
In absence of of driving $\lambda\to 0$ one can check with $\mu_1\to 1$ (as $\mathcal{J}_0\left(\frac{2\lambda}{\Omega}\right)\to 1$ for $\lambda \rightarrow 0$) that this falls back to the exact pure-dephasing solution~\cite{Lidar_2001,leggett1987a}
with {constant populations~\eqref{EQ:pdp_pop} and decaying coherences~\eqref{EQ:pdp_exact}}.

%
Furthermore, at vanishing system-reservoir coupling we have $\Sigma(t) \to 1$, where one can immediately verify that the purity of an initial state is preserved.
Finally, we can see that for reasonable spectral densities we find that $\lim_{t\to\infty} \Sigma(t)=0$, such that the long-term asymptotics of~\eqref{EQ:fastdriving} is given by
\begin{align}
    \expval{\sigma^y} &\to -\sin\left(\frac{2\lambda}{\Omega} \sin(\Omega t)\right) \expval{\sigma^z}_0\,,\nn
    \expval{\sigma^z} &\to +\cos\left(\frac{2\lambda}{\Omega} \sin(\Omega t)\right) \expval{\sigma^z}_0\,.
\end{align}
Importantly, we stress that the fast-driving solution is valid also for stronger system-reservoir couplings~\cite{restrepo2016a}.
{One can extend this perturbative scheme by transforming Eq.~\eqref{EQ:transformed_fastdriving} into yet another picture and perform the RWA then in this frame. 
Doing so has for the parameters considered not led to significant changes, which makes us believe that the above fast-driving solution can for large $\Omega$ indeed be considered a benchmark solution.}

\subsection{Master equation solutions}

Since we are interested in the fast driving regime here, we consider kick operator $U_{\rm kick}(t)\approx \exp\{-\ii \lambda/\Omega \sin(\Omega t) \sigma^x\}$ and system Floquet Hamiltonian 
$\bar{H}\approx\frac{\Delta}{2} {\cal J}_0\left(\frac{2\lambda}{\Omega}\right) \sigma^z$
also by applying the RWA in the transformed frame.
Considering the scaling of the Bessel functions at small arguments, we therefore consistently only keep the lowest order terms with $n\in\{-1,0,+1\}$ in Eq.~(\ref{EQ:coupling_op}).
Then, the coupling operator in the interaction picture can be approximately written as
\begin{align}
    \f{A}(t) &\approx
    {\cal J}_0\left(\frac{2\lambda}{\Omega}\right) \sigma^z + 2 \sin(\Omega t) {\cal J}_1\left(\frac{2\lambda}{\Omega}\right) e^{+\ii \bar{H} t} \sigma^y e^{-\ii \bar{H} t}\nn
    &\qquad+\ord\left\{\frac{\lambda^2}{\Omega^2}\right\}\,,
\end{align}
which suffices to set up all master equations.
We have numerically checked convergence in the shown parameter regime by considering also the case $n\in\{-2,\ldots,+2\}$ (not shown).

\subsection{Comparison}

\begin{figure}
\centering
\includegraphics[width=0.45\textwidth,clip=true]{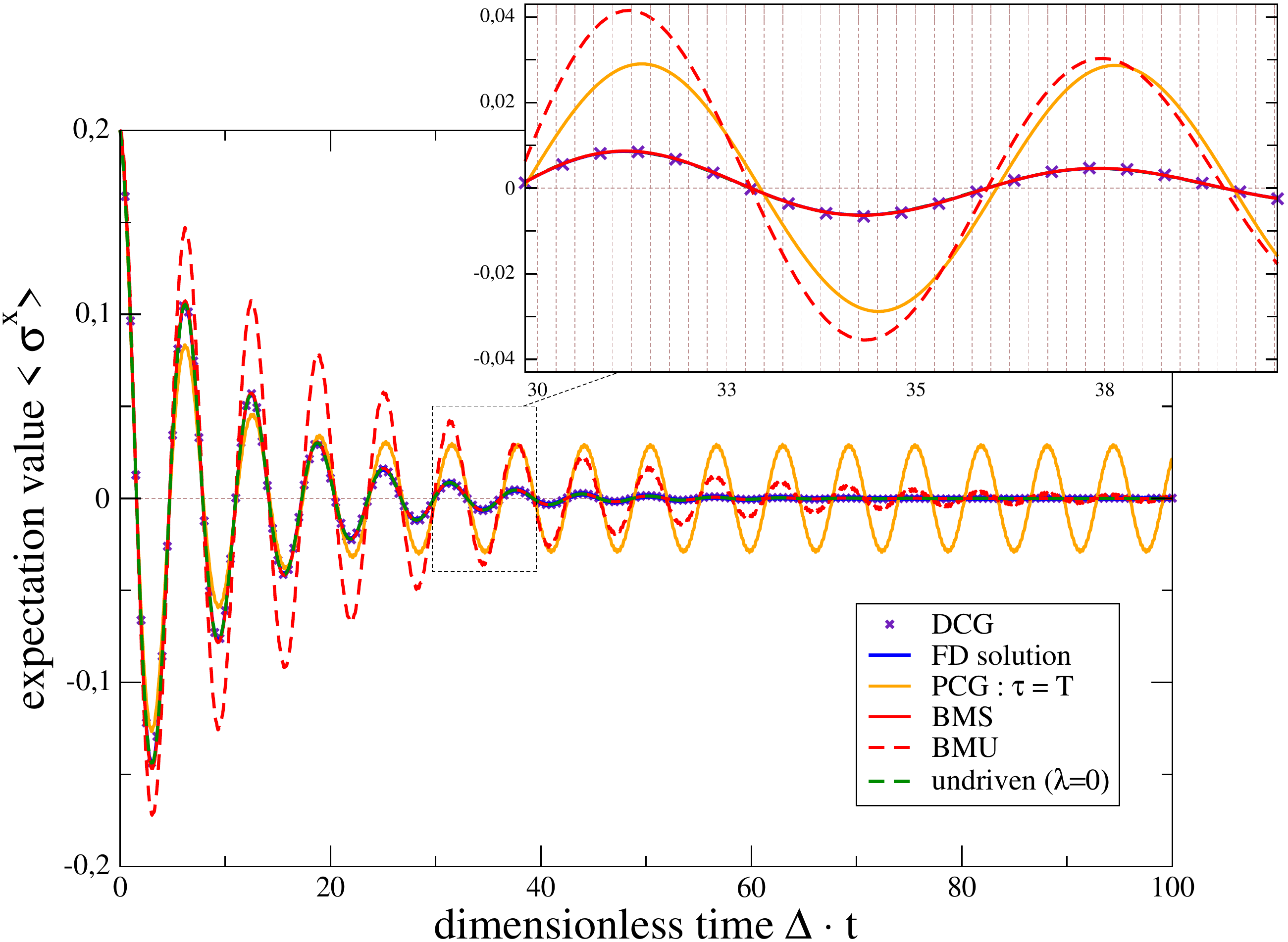}
\caption{(Color Online) Plot of  the  time-dependent  expectation  value  of  Pauli matrix $\sigma^x$ comparing the {no-driving solution (dashed green)}, the analytical fast-driving solution (blue), PCG (orange), BMS (solid red), BMU (dashed red) and DCG (purple cross). The gridlines mark multiples of the driving period $T$. A higher order approximation for the driving yields similar results (not shown). Parameters: $\Omega=25 \; \Delta$, $\lambda=\frac{1}{2}\; \Delta$, $\beta \, \Delta=1$, $\Gamma_0=0.05$, $\omega_c=15 \; \Delta$, $\rho_0=1/2(\f{1}+0.2 \sigma^x+0.4 \sigma^z)$. 
}
\label{fig_seb_sx}
\end{figure}

\begin{figure}
\includegraphics[width=0.45\textwidth,clip=true]{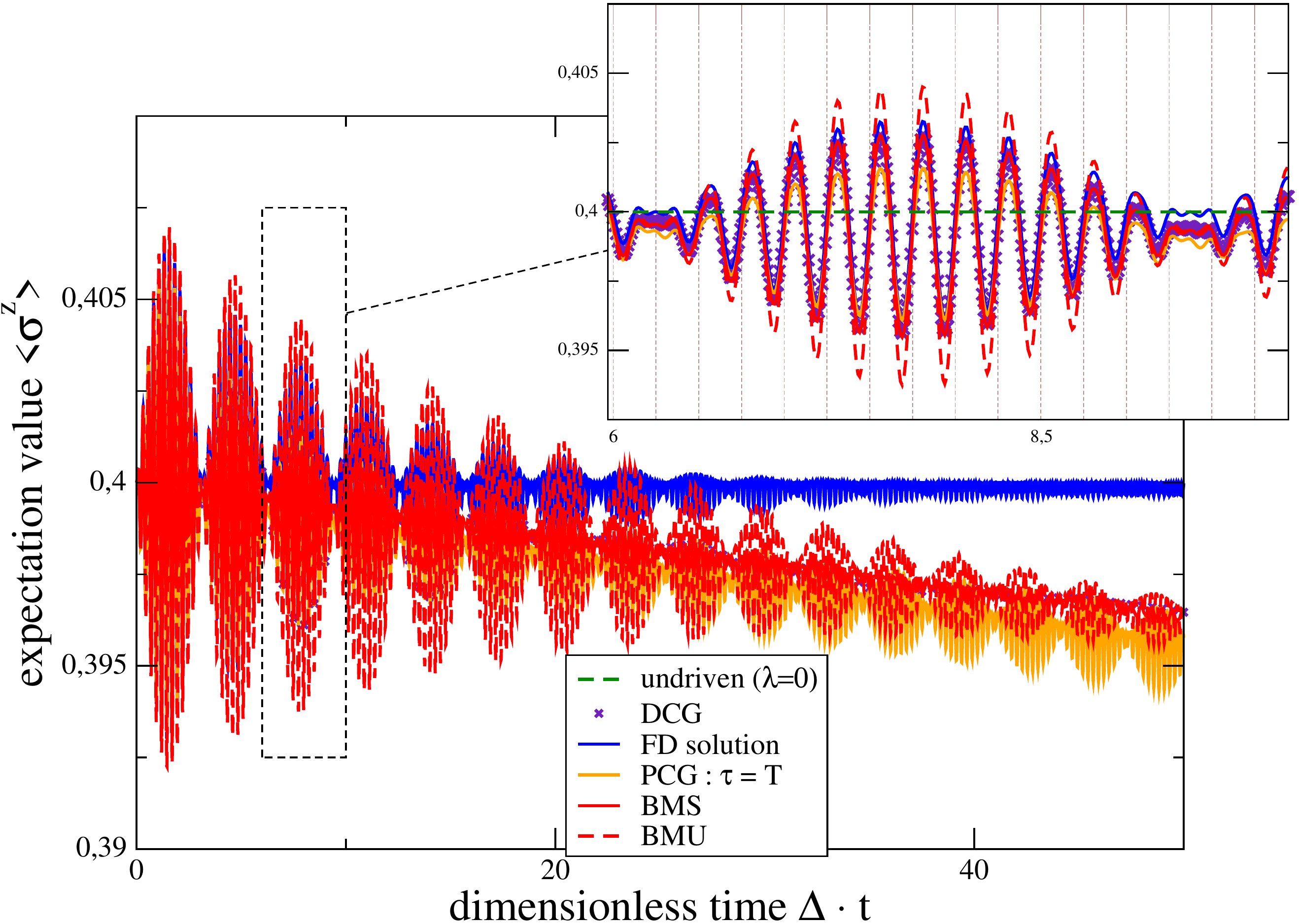}
\caption{(Color Online) Analogous to Fig.~\ref{fig_seb_sx}, but for $\expval{\sigma^z}$. Whereas for the transient dynamics DCG, PCG, and BMS solutions are close to the benchmark fast driving solution (inset), significant differences arise for longer times (main plot).}
\label{fig_seb_sz}
\end{figure}

In the limit of both fast driving and weak coupling we can compare the benchmark solution with the perturbative {PCG}/DCG/BMS/BMU approaches.
{The expectation value $\expval{\sigma^x}$ is shown in Fig.~\ref{fig_seb_sx}.}
{
One can see that only the BMU (dashed red) and PCG (orange) solutions fail to capture the correct dynamics, whereas BMS (solid red) and DCG (symbols) solutions closely approximate it.
It should be kept in mind however that for such fast drivings, already the naive RWA in the original picture -- yielding the solution~\eqref{EQ:pdp_exact} with $\lambda=0$ (dashed green) -- would yield acceptable results for $\expval{\sigma^x}$.
This is different for the expectation value $\expval{\sigma^z}$, which we show in Fig.~\ref{fig_seb_sz}.
}
For this expectation value, a naive RWA treatment would just predict a constant evolution -- which is indeed best captured by the long-term fast-driving evolution and not by the perturbative approaches.

In both figures we observe that the DCG approach is closest to the FD solution. 
It has however the disadvantage that the generator has to be re-computed for any desired time.
Still, all perturbative approaches fail to correctly capture the long-term dynamics of the fast-driving solution~\eqref{EQ:fastdriving}, which is not surprising as the latter is non-perturbative in the system-reservoir coupling strength.

\begin{figure}
\includegraphics[width=0.45\textwidth,clip=true]{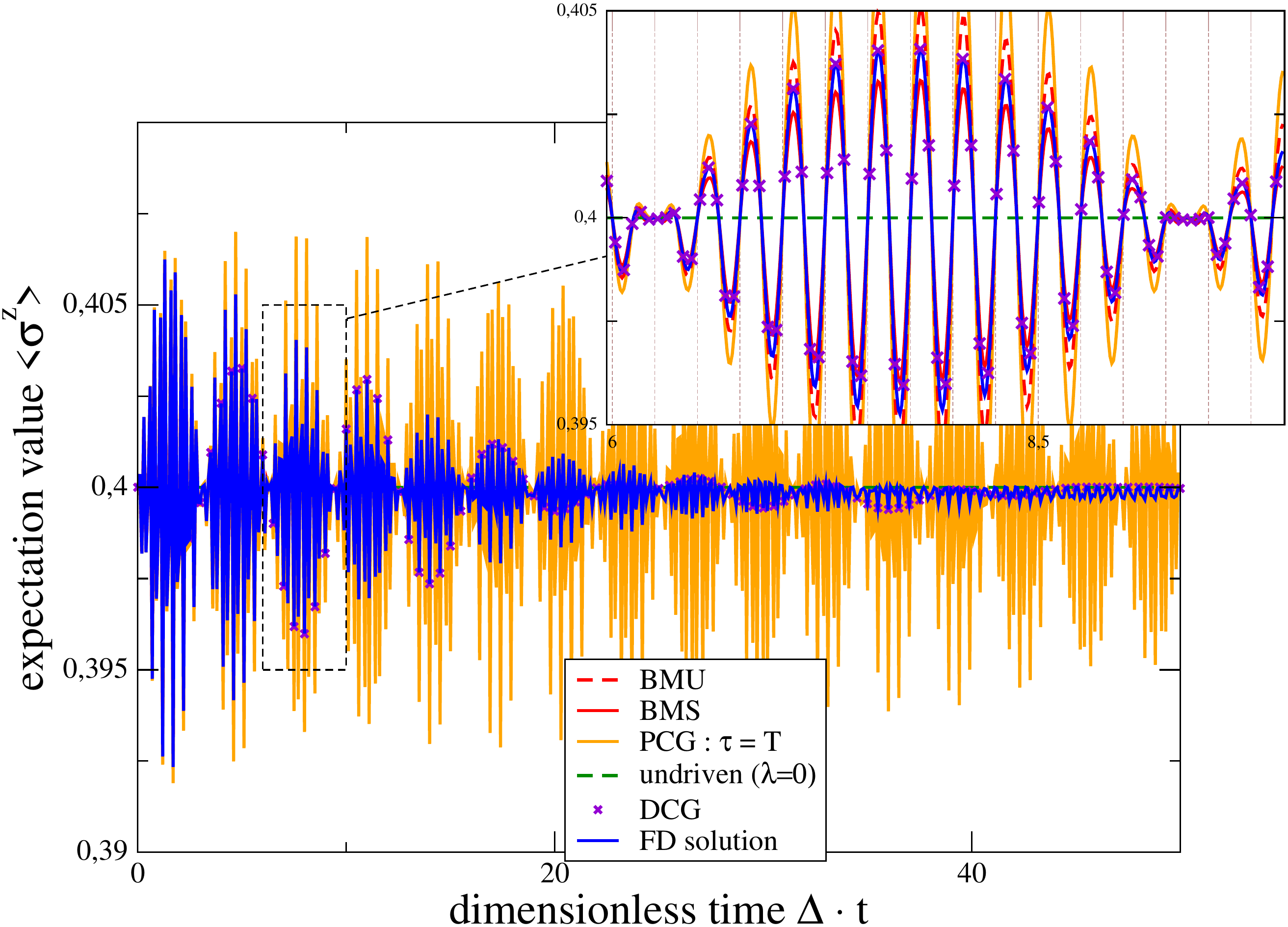}
\caption{(Color Online) Analogous to Fig.~\ref{fig_seb_sz}, but for smaller cutoffs implying a slower decay of the reservoir correlation function $\omega_c = \Delta$. 
While DCG and fast-driving benchmark still agree perfectly, the Markovian approaches (BMS, BMU, and PCG) all deviate strongly.
}
\label{fig_seb_sz_b}
\end{figure}

So far, we have considered the regime $\Omega \gg \omega_c \gg \Delta$, which corresponds to $\tau_{\rm drv} \ll \tau_{\rm dec} \ll \tau_{\rm int}$.
We see that BMS and BMU solution fail to describe the long-term dynamics correctly, but this is also the case for the DCG approach, which captures the short-term dynamics by construction.
In the regime $\Omega \gg \omega_c \approx \Delta$, the BMS, BMU, and PCG solutions perform significantly worse, since then the Markovian approximation is not valid anymore.
In contrast, the DCG solution still matches the fast-driving benchmark very well, as it is not subject to a Markovian approximation.
We show this in Fig.~\ref{fig_seb_sz_b} for the expectation value $\expval{\sigma^z}$ (the expectation value of $\expval{\sigma^x}$ looks similar, but there the driving has little effect anyways).


\section{Conclusions}

Our intention in this study was to find an improved time-independent Lindblad (Markovian~\cite{breuer2012a}) master equation applicable to periodically driven systems. 
For this, we analyzed driven qubit systems coupled to thermal reservoirs with coarse-graining approaches.
All methods used had the formal advantage of being in LGKS form, which unconditionally preserves all density matrix properties and leads to a stable numerical propagation.
The negative result is that the analyzed coarse-graining approach over one period of the driving (PCG) did not match the expectations, i.e., it was inferior to the DCG approach, even in the special case of pure dephasing.

On the positive side, we found that the BMS approach can yield quite reliable results in the long term limit provided the secular approximation is performed in a proper way. 
The DCG approach performed at least as good as the BMS variant in the long-term limit and approaches the exact solution by construction in the short-term limit.
As the corresponding systematic expansion converges in the weak-coupling regime (see App.~\ref{APP:convergence}), we consider it as the most accurate second-order perturbative solution among the schemes tested.
Its drawback is the computational cost, since for each time a suitable dissipator needs to be calculated numerically.
To avoid this, we note that by using spectral densities with a simpler polynomial structure, the involved integrals in~(\ref{EQ:CGx}) can be performed analytically.
The fact that it is not possible to capture the full dynamics faithfully with a single coarse-graining time could be taken as a hint that the restriction to LGKS dynamics is too severe.
Using a Redfield approach however may be highly unstable for driven systems even in the weak-coupling limit, such that one should rather aim at deriving Kraus representations from first principles.

\begin{acknowledgments}
R.H. acknowledges financial support by the Japanese Student Services Organization JASSO.
The authors gratefully acknowledge discussions with J. Abla{\ss}mayer, T. Becker, A. Eckardt, and A. Schnell.
\end{acknowledgments}

\appendix


\section{Coarse-graining generators}

\subsection{General coarse-graining derivation}\label{app:details}

{
To evaluate the r.h.s. of the defining equation~\eqref{EQ:cgdef}, we first calculate the time evolution operator in the interaction picture, where it follows the equation
\begin{align}
    \frac{d}{dt} \f{U}(t,t_0)= -\ii \f{H}_I(t,t_0) \f{U}(t,t_0)\,,
\end{align}
together with the initial condition $\f{U}(t_0,t_0) =  \f{1}$.
Formally integrating this equation and repeatedly inserting the solution into the r.h.s. yields the representation
\begin{align}\label{EQ:timeevolution}
    \f{U}(t,t_0) &= \sum_{n=0}^\infty (-\ii)^n \int\limits_{t_0}^t dt_1 \int\limits_{t_0}^{t_1} dt_2 \ldots \int\limits_{t_0}^{t_{n-1}} dt_n\nn 
    &\qquad\times\f{H}_I(t_1) \f{H}_I(t_2) \ldots \f{H}_I(t_n)\,,
\end{align}
where the nested integration bounds make the time-ordering explicit and the $n=0$-term is understood as identity.
Now, inserting a generic tensor product decomposition of the interaction Hamiltonian
\begin{align}
    \f{H}_I(t,t_0) &= \sum_\alpha \f{A}_\alpha(t,t_0) \otimes \f{B}_\alpha(t,t_0)\nn
    &= \sum_\alpha U_S^\dagger(t,t_0) A_\alpha U_S(t,t_0)\nn
    &\qquad\otimes e^{+\ii H_B (t-t_0)} B_\alpha e^{-\ii H_B (t-t_0)}
\end{align}
for system and bath coupling operators $A_\alpha$ and $B_\alpha$~\footnote{Our derivation also allows for driven system and bath coupling operators that are time-dependent already in the Schr{\"o}dinger picture, but we will drop this for brevity.}, respectively, 
we see that the isolated time evolution operator of the (undriven) reservoir can be given explicitly, whereas the explicit calculation is required for the isolated time evolution operator of the (driven) system $\f{U}_S(t,t_0)$.

Applied to the periodically-driven Schr\"odinger equation, Floquet theory predicts that the time evolution operator can be expressed as a product of a unitary and periodic kick operator (that has the period of the driving) and the exponential of a Floquet Hamiltonian~\cite{eckardt2015a}.
Specifically, for $t_0=0$ one may write 
\begin{align}
    U_S(t)&=U_{\rm kick}(t)e^{-\ii\bar{H}t}\,,\\
    U_{\rm kick}(t)&=U_{\rm kick}(t+T)\,,\qquad
    U_{\rm kick}(nT)=\mathds{1}\,,\nonumber
      \label{floquet zerlegung}
\end{align}
with $n \in \mathds{Z}$, such that the stroboscopic evolution between full periods is given by the Floquet Hamiltonian
$U_S(nT)=e^{-\ii\bar{H}n T}$.
It should be stressed here that even in this simple case the above decomposition is not unique.
For example, introducing the eigenbasis of the Floquet Hamiltonian via $\bar{H}\ket{\bar{a}} = \bar{E}_a \ket{\bar{a}}$, we can see that the time evolution operator remains invariant under the simultaneous transformations $\bar{H}\to\bar{H}+m \Omega \ket{\bar{a}}\bra{\bar{a}}$ and $U_{\rm kick}(t)\to U_{\rm kick}(t) e^{+\ii m \Omega t \ket{\bar{a}}\bra{\bar{a}}}$, which maintains the periodicity of the kick operator.
This gauge degree of freedom allows one to deliberately choose the convention that all eigenenergies of the Floquet Hamiltonian should lie in the first Brillouin zone $\bar{E}_a \in [-\Omega/2,+\Omega/2)$.
Hence, the system coupling operators in the interaction picture $\f{A}_\alpha(t)=\f{A}_\alpha(t,0)$ can be expressed as:
\begin{align}
     \boldsymbol{A}_\alpha(t)&=U_S^{\dagger}(t)A_\alpha U_S(t)=  e^{\ii\bar{H}t}U_{\rm kick}^{\dagger}(t)A_\alpha U_{\rm kick}(t)e^{-\ii\bar{H}t}\nn
    & =\sum_{n=-\infty}^{+\infty}e^{\ii\bar{H}t}\hat{A}_\alpha^n e^{\ii n\Omega t}e^{-\ii\bar{H}t}\,, 
\end{align}
where due to the periodicity of the kick operator we have expanded 
$U_{\rm kick}^{\dagger}(t)A_\alpha U_{\rm kick}(t)$ in a Fourier series 
\begin{align}
 \hat{A}_\alpha^n=\frac{\Omega}{2 \pi}
 \int\limits^{T/2}_{-T/2}U_{\rm kick}^{\dagger}(t)A_\alpha U_{\rm kick}(t)e^{-\ii n\Omega t} dt, && T=\frac{2 \pi}{\Omega}\,.
\end{align}
Now, in the eigenbasis of the Floquet Hamiltonian we can also write this as
\begin{align}\label{EQ:timedep_system}
    \boldsymbol{A}_\alpha(t_1) &= \sum_{ab}\sum_n \bra{\bar{a}}\hat{A}_\alpha^n \ket{\bar{b}}e^{\ii(\bar{E}_a-\bar{E}_b+n\Omega)t_1}\ket{\bar{a}}\bra{\bar{b}}\nn
    &\equiv \sum_{ab} \sum_n  A_{\alpha,ab}^{+n} e^{+\ii (\bar{E}_a-\bar{E}_b + n \Omega) t_1} L_{ab}\,,\nn
    \boldsymbol{A}_{\bar\alpha}(t_2) &= \sum_{cd}\sum_{n'} \bra{\bar{c}}\hat{A}_{\bar\alpha}^{n'} \ket{\bar{d}} e^{\ii(\bar{E}_c-\bar{E}_d+n'\Omega)t_2}\ket{\bar{c}}\bra{\bar{d}}\nn
    &\equiv \sum_{cd} \sum_{n'}  A_{\bar\alpha,dc}^{-n'} e^{-\ii (\bar{E}_c-\bar{E}_d + n' \Omega) t_2} L_{dc}\,,
\end{align}
where $L_{ab} \equiv \ket{\bar{a}}\bra{\bar{b}}$ and we exchanged $c \leftrightarrow d$ and $n'\to-n'$ in the last line.

The reservoir contributions to the full time evolution operator in the interaction picture can be re-organized into reservoir correlation functions, which for a reservoir at equilibrium depend only on the difference of time arguments.
Therefore, we can introduce their even and odd Fourier transforms~\footnote{
We note that higher-order correlation functions can be related to two-point correlation functions for linear operators and thermal harmonic reservoirs.}
\begin{align}\label{EQ:corrfuncdef}
    C_{\alpha \bar{\alpha}}(t_1-t_2)&=\frac{1}{2 \pi}\int d\omega \gamma_{\alpha\bar{\alpha}}(\omega) e^{-\ii \omega (t_1-t_2)}\,,\nn
     C_{\alpha \bar{\alpha}}(t_1-t_2)\sgn (t_1-t_2)&=\frac{1}{2 \pi}\int d\omega \sigma_{\alpha\bar{\alpha}}(\omega) e^{-\ii \omega (t_1-t_2)}\,.
\end{align}

Inserting both the system and reservoir contributions above into~\eqref{EQ:cgdef}, expanding also its l.h.s. to first order and solving for $\f{{\cal L}_\tau}$, the temporal integrals can be analytically performed with the help of
\begin{align}
f_{t_0}^\tau(\alpha,\beta,\omega)&\equiv
  \frac{1}{2 \pi \tau}\int\limits_{t_0}^{t_0+\tau}\int\limits_{t_0}^{t_0+\tau}e^{-\ii\omega(t_1-t_2)}e^{\ii\alpha t_1}e^{-\ii\beta t_2}dt_1 dt_2\nn
  &= \frac{\tau}{2 \pi}e^{\ii(\alpha-\beta)t_0} 
  e^{\ii(\alpha-\beta)\tau/2}\sinc \left[(\alpha-\omega)\frac{\tau}{2}\right]\nn
  &\qquad\times \sinc \left[(\beta-\omega)\frac{\tau}{2}\right]\,.
    \label{time int sinc}
\end{align}
These definitions are used in the generators in Eqn.~\eqref{EQ:CGx} in the main text and in Eqns.~\eqref{EQ:CGx_lt},~\eqref{EQ:bms}, and~\eqref{EQ:bmu} below.

The function defined above has additional useful properties. 
For example, for discrete $\alpha$ and $\beta$ one finds
\begin{align}\label{EQ:deltafunc}
    \lim_{\tau\to\infty} f_{t_0}^\tau(\alpha,\beta,\omega) = \delta_{\alpha\beta} \delta(\alpha-\omega)\,,
\end{align}
which is used to derive the long-term limit of the coarse-graining generator in App.~\ref{APP:longterm}.
Additionally, also for finite $\tau$, integrals involving this function remain bounded as we show below.

\subsection{Convergence of the expansion}\label{APP:convergence}

Inserting the expansion for the time-evolution operator~\eqref{EQ:timeevolution} into the defining coarse-graining equation~\eqref{EQ:cgdef}, we additionally obtain many temporal integrals over higher-order correlation functions -- some of them equipped with Heaviside-$\Theta$ functions depending on time differences that enforce the time-ordering.
For linear couplings and harmonic reservoirs, only correlation functions with an even number of operators remain, and furthermore, these can then be expressed as products of two-point correlation functions.
Thus, defining Fourier transforms as before, we also obtain for higher-order terms the nascent $\delta$-function~\eqref{time int sinc} together with a function $g(\omega)$ that contains Fourier transforms in analogy to $\gamma_{\alpha\bar\alpha}(\omega)$ and $\sigma_{\alpha\bar\alpha}(\omega)$ from Eq.~\eqref{EQ:corrfuncdef}.
We can bound integrals over products of this nascent $\delta$-function with other bounded functions $g(\omega)$ by the essential supremum of the latter
\begin{align}
 D &\equiv \abs{\int d\omega g(\omega) f_{t_0}^\tau(\omega_\alpha,\omega_\beta,\omega)}\nn
  & \le \int d\omega \abs{g(\omega) f_{t_0}^\tau(\omega_\alpha,\omega_\beta,\omega)}\nn
  & \le \left[{\rm ess\;sup}_\omega \abs{g(\omega)}\right] 
  \int \abs{f_{t_0}^\tau(\omega_\alpha,\omega_\beta,\omega)} d\omega\nn
  & \le \left[{\rm ess\;sup}_\omega \abs{g(\omega)}\right] \sqrt{\int\frac{\tau}{2\pi} {\rm sinc}^2\left[(\omega_\alpha-\omega)\frac{\tau}{2}\right] d\omega}\times\nn
  &\qquad\times \sqrt{\int\frac{\tau}{2\pi} {\rm sinc}^2\left[(\omega_\beta-\omega)\frac{\tau}{2}\right] d\omega}\nn
  &= {\rm ess\;sup}_\omega \abs{g(\omega)}\,.
\end{align}
Here, in the second inequality we have used H\"olders inequality for the $1$-norm and the $\infty$-norm, and in the third inequality we have first inserted~\eqref{time int sinc} and then used H\"olders inequality for the $2$-norm.
{Thus, if the maximum of even and odd Fourier transforms of the correlation functions (that enter as $g(\omega)$) is small, i.e., in the weak-coupling limit, we may expect convergence of the series for all $\tau$.}

For the bosonic reservoirs that we consider, a finite Fourier transform of the reservoir correlation function requires at least ohmic spectral densities.
If, for example, the Fourier transform of the reservoir correlation function is given by $\gamma(\omega) = \Gamma_0 \omega e^{-\abs{\omega}/\omega_c} [1+1/(e^{\beta\omega}-1)]$, we find at zero temperatures $\gamma(\omega) \le \Gamma_0 \omega_{\rm c} e^{-1}$ and at high temperatures $\gamma(\omega)\le \Gamma_0/\beta$.
}

\subsection{Dynamical coarse-graining (DCG)}\label{APP:dcg}

For dynamical coarse-graining, the time-dependent system density matrix is expressed as
\begin{align}\label{EQ:dcg}
    \f{\rho_S}(t) = \exp\left\{\f{{\cal L}_t}\cdot t\right\} \f{\rho_S}(t_0)\,.
\end{align}
This density matrix then formally evolves according to 
$\frac{d}{dt} \f{\rho_S} = \left[\left(\frac{d}{dt}e^{\f{{\cal L}_t}\cdot t}\right)e^{-\f{{\cal L}_t}\cdot t}\right] \f{\rho_S}(t)$, where the term in square brackets can be seen as a non-Markovian generator of the dynamical coarse-graining evolution~\cite{rivas2017a}.
The full term in square brackets is not of LGKS form.
Nevertheless, the scheme preserves all density matrix properties, which follows as $\f{{\cal L}_t}$ is of LGKS form for fixed $t$, such that the scheme~\eqref{EQ:dcg} is an interpolation through physically valid density matrices.

\subsection{Initial-period coarse-graining (PCG)}\label{APP:cg1}

Coarse-graining only over the initial period ($t_0=0$ and $\tau=T$), the PCG-coarse-graining generator is obtained from~\eqref{EQ:CGx}
by inserting 
\begin{align}
f_0^T(\omega_1,\omega_1,\omega) 
&= \frac{T}{2\pi} e^{\ii (\omega_1-\omega_2) T/2}\\
&\qquad\times\sinc\left[(\omega_1-\omega)\frac{T}{2}\right]\sinc\left[(\omega_2-\omega)\frac{T}{2}\right]\,.\nonumber
\end{align}
The generator then still contains an integration over $\omega$, which -- depending on the form of the spectral density -- may need to be solved numerically.

\subsection{Long-term coarse-graining}\label{APP:longterm}

When $\tau\to\infty$, we get from Eq.~\eqref{EQ:deltafunc} that the remaining integral in~\eqref{EQ:CGx} can be fully resolved, leading to
\begin{align}
\label{EQ:CGx_lt}
    {\bf H_{\rm LS}^\infty} &= 
     \frac{1}{2\ii} \sum_{\alpha\bar{\alpha}} \sum_{nn'}\sum_{abcd}
     \sigma_{\alpha\bar{\alpha}}(\bar{E}_a-\bar{E}_b+n\Omega)\nn    
    & \qquad \times
    \delta_{\bar{E}_a-\bar{E}_b+n \Omega,\bar{E}_c-\bar{E}_d+n' \Omega}
    A_{\alpha,ab}^{n} A_{\bar{\alpha},dc}^{-n'}
    {L_{ab} L_{dc}}\,,\nn
\mathcal{D}^\infty \boldsymbol{\rho_S} &= \sum_{\alpha\bar{\alpha}} \sum_{n n'}\sum_{abcd} 
\gamma_{\alpha\bar{\alpha}} (\bar{E}_a-\bar{E}_b+n \Omega)\nn
&\qquad\times\delta_{\bar{E}_a-\bar{E}_b+n \Omega,\bar{E}_c-\bar{E}_d+n' \Omega} A_{\alpha,ab}^{+n} A_{\bar\alpha,dc}^{-n'}\nn
    & \qquad \times
    {
    \left[ L_{dc} \boldsymbol{\rho_S} L_{ab} -\frac{1}{2}\left\{L_{ab} L_{dc} ,\boldsymbol{\rho_S}\right\}\right]
    }\,.
\end{align}
Thus, the corresponding generator can be readily evaluated, and in particular the long-term results from App.~\ref{APP:dcg} will coincide with the long-term results from the equation above.

\subsection{Floquet-BMS master equation}\label{APP:bms}

{In Eq.~\eqref{EQ:CGx_lt},} the evaluation of the resonance condition resulting from the Kronecker-$\delta$
\begin{align}
    \bar{E}_a-\bar{E}_b+n \Omega = \bar{E}_c-\bar{E}_d + n' \Omega
\end{align}
may require some care.
The typical argument is that for fast driving, the above resonance can only be met if separately $n'=n$ and $\bar{E}_a-\bar{E}_b=\bar{E}_c-\bar{E}_d$.
However, the applicability of this argument critically depends on the Floquet spectrum. 
If for a two-level system we by chance have Floquet energies well within the first Brillouin zone $\bar{E}_a\in\{-\Omega/4,+\Omega/4\}$, this generates energy differences $\bar{E}_a-\bar{E}_b\in\{-\Omega/2,0,+\Omega/2\}$.
In this case, the above resonance condition can also be met with $\bar{E}_a-\bar{E}_b=+\Omega/2$, $\bar{E}_c-\bar{E}_d=-\Omega/2$ and $n'=n+1$, demonstrating that in general, the long-term limit of periodically driven coarse-graining master equation need not coincide with the significantly simpler Floquet-BMS master equation that results from demanding the resonance separately (i.e., by setting $n'=n$ and $\bar{E}_a-\bar{E}_b = \bar{E}_c-\bar{E}_d$)
\begin{align}\label{EQ:bms}
    \f{H_{\rm LS}^{\rm BMS}} &=
     \frac{1}{2\ii} \sum_{\alpha\bar{\alpha}} \sum_{n}\sum_{abcd}
     \sigma_{\alpha\bar{\alpha}}(\bar{E}_a-\bar{E}_b+n\Omega)\nn    
    & \qquad \times
    \delta_{\bar{E}_a-\bar{E}_b,\bar{E}_c-\bar{E}_d}
    A_{\alpha,ab}^{n} A_{\bar{\alpha},dc}^{-n}
    {L_{ab} L_{dc}}\,,\nn
\mathcal{D}^{\rm BMS} \boldsymbol{\rho_S} &= \sum_{\alpha\bar{\alpha}} \sum_{n}\sum_{abcd} 
\gamma_{\alpha\bar{\alpha}} (\bar{E}_a-\bar{E}_b+n \Omega)\nn
&\qquad\times\delta_{\bar{E}_a-\bar{E}_b,\bar{E}_c-\bar{E}_d} A_{\alpha,ab}^{+n} A_{\bar\alpha,dc}^{-n}\nn
    & \qquad \times
    {
    \left[ L_{dc} \boldsymbol{\rho_S} L_{ab} -\frac{1}{2}\left\{L_{ab} L_{dc} ,\boldsymbol{\rho_S}\right\}\right]
    }\,.
\end{align}

\subsection{The ultra-secular limit (BMU)}\label{APP:bmu}

Even when~\eqref{EQ:bms} is applicable, we note that the remaining Kronecker-$\delta$ leaves many terms that are often neglected. 
For example, the above resonance can always be trivially fulfilled with $\bar{E}_a=\bar{E}_b$ and $\bar{E}_c=\bar{E}_d$, even for the undriven case.
For many models, one has that $\gamma_{\alpha\bar\alpha}(0)\to 0$, such that such terms would not contribute anyways for undriven systems, where $\Omega\to 0$.
But in general they will have to be kept. 
For comparison we therefore also state the ultra-secular approximation (BMU), where only terms with $a=c$ and $b=d$ are kept 
\begin{align}\label{EQ:bmu}
{\bf H_{\rm LS}^{\rm BMU}} &= 
     \frac{1}{2\ii} \sum_{\alpha\bar{\alpha}} \sum_{n}\sum_{ab}
     \sigma_{\alpha\bar{\alpha}}(\bar{E}_a-\bar{E}_b+n\Omega)\nn
     &\quad\times
     A_{\alpha,ab}^{n} A_{\bar{\alpha},ba}^{-n}
     {L_{ab} L_{ba}}\,,\nn
\mathcal{D}^{\rm BMU}\boldsymbol{\rho_S} &= \sum_{\alpha\bar{\alpha}} \sum_{n}\sum_{ab} 
\gamma_{\alpha\bar{\alpha}} (\bar{E}_a-\bar{E}_b+n \Omega)
A_{\alpha,ab}^{+n} A_{\bar\alpha,ba}^{-n}\nn
    & \quad \times\left[ L_{ba} \boldsymbol{\rho_S} L_{ab} -\frac{1}{2}\left\{L_{ab} L_{ba} ,\boldsymbol{\rho_S}\right\}\right]\,.
\end{align}


\section{Exact pure-dephasing solution}\label{APP:pd}

\subsection{General solution}

{For a driven Hamiltonian of pure dephasing form~\eqref{EQ:pure_dephasing}, the exact dynamics} can be calculated by using a polaron (or Lang-Firsov~\cite{lang1963a}) transform:
\begin{align}
    U_{\rm p}=e^{A \otimes \left(\frac{h_k}{\omega_k} b_k - \frac{h_k^*}{\omega_k} b_k^\dagger\right)}.
    \label{polaron trafo}
\end{align}
Applying this  transformation to system and bath operators yields
\begin{align}
U_{\rm p}^{\dagger} H_S(t) U_{\rm p} &=  H_S(t)\,,\qquad
U_{\rm p}^{\dagger} A U_{\rm p}=A\,,\\
U_{\rm p}^{\dagger} b_k U_{\rm p}&= b_k - \frac{h^*_k}{\omega_k}A\,,\qquad
U_{\rm p}^{\dagger} b_k^{\dagger} U_{\rm p}=b_k^{\dagger}-\frac{h_k}{\omega_k}A\,,\nonumber
\end{align}
which leads to decoupled system and bath Hamiltonians
\begin{align}
U_{\rm p}^{\dagger} H(t) U_{\rm p} &=H_S(t) -\left( \sum_k \frac{|h_k|^2}{\omega_k}\right)A^2+\sum_k\omega_k b_k^{\dagger} b_k\nn
&\equiv\bar{H}_S(t)+\sum_k\omega_k b_k^{\dagger} b_k,
\label{decoupled sb}
\end{align}
with $\bar{H}_S(t)$ denoting an effective system Hamiltonian. Hence, the time evolution operator in the polaron picture $ \tilde{U}(t)$ is given by the system and bath evolution separately
\begin{align}
    \tilde{U}(t)=\mathcal{T}\left\{e^{-\ii \int_0^t \bar{H}_S(t') dt'}\right\} e^ { -\ii \sum_k\omega_k b_k^{\dagger} b_k t}\,,
\end{align}
where $\mathcal{T}$ denotes the time-ordering operator of the system.
For a general unitary transformation $V(t)$ into another frame via $\ket{\Psi(t)} = V(t) \ket{\tilde\Psi(t)}$, the time evolution operator in the original frame $U(t)$ can be expressed in terms of the time evolution operator in the new frame $\tilde{U}(t)$ as $U(t)=V(t) \tilde{U}(t) V^\dagger(0)$.
Thus, for the time-independent polaron transform this relation yields
\begin{align}
    U(t)=U_{\rm p} \tilde{U}(t) U_{\rm p}^\dagger.
\end{align}
From this, the expectation value of any system observable $O_S$,
\begin{align}
    \expval{O_S} &= \trace{(O_S \otimes \f{1}_B) \rho(t)}\nn
    &=\trace {U^\dagger(t)(O_S \otimes \f{1}_B) U(t)\rho^0},
\end{align}
can be calculated as:
\begin{align}
    \expval{O_S} =\trace{ U_{\rm p} \tilde{U}^{\dagger}(t) U_{\rm p}^{\dagger} (O_S \otimes \f{1}_B) U_{\rm p} \tilde{U}(t) U_{\rm p}^{\dagger} \rho^0}\,.
\end{align}

\subsection{Two-level system}

For the example of a two level system,
\begin{align}
    H_S(t)= \frac{\Delta}{2} \sigma^z{+\lambda \cos\Omega t}\,,\qquad
    A=\sigma^z,
\end{align}
the populations $\rho_{S,00}(t)=\frac{1}{2}\langle \mathds{1}+\sigma^z\rangle$ and $\rho_{S,11}(t)=\frac{1}{2}\langle \mathds{1}-\sigma^z\rangle$ are constant as  $\mathds{1}$ and $\sigma^z$ commute with $U_{\rm p}$ and $\tilde{U}(t)$, 
{and we obtain~\eqref{EQ:pdp_pop} in the main text.}

For the coherences $\rho_{01}=\expval{\sigma^-}$ and $\rho_{10}=\expval{\sigma^+}$, the calculations are more involved. Using that
\begin{align}
U_{\rm p}^{\dagger} \sigma^{\pm} U_{\rm p}=e^{\pm2 \sum_k\left(\frac{h_k^*}{\omega_k}b_k^{\dagger}-\frac{h_k}{\omega_k}b_k \right)}   \sigma^{\pm},
\end{align}
and organising bath and system parts together, one gets
\begin{align}
  \expval{\sigma^\pm} &={\rm Tr} \Big\{U_{\rm p} U_{B}^\dagger(t) e^{\pm2 \sum_k\left(\frac{h_k^*}{\omega_k}b_k^{\dagger}-\frac{h_k}{\omega_k}b_k \right)}  U_{B}(t)U_S^\dagger(t) \sigma^{\pm}
  \nn & \qquad \times U_S(t) U_{\rm p}^{\dagger} \rho^0 \Big\}, 
\end{align}
with $U_{B}(t)=e^{-\ii \sum_k\omega_k b_k^{\dagger} b_k t}$ and $U_S(t)=e^{-\ii\left(\frac{\Delta}{2}t+\frac{\lambda}{\Omega}\sin(\Omega t)\right)\sigma^z}$. 
Now, one can use that
\begin{align}
  &e^{\ii \sum_k\omega_k b_k^{\dagger} b_k t} e^{\pm2 \sum_k\left(\frac{h_k^*}{\omega_k}b_k^{\dagger}-\frac{h_k}{\omega_k}b_k \right)} e^{-\ii \sum_k\omega_k b_k^{\dagger} b_k t}\nn
  &\qquad =e^{\pm2 \sum_k\left(\frac{h_k^*}{\omega_k}b_k^{\dagger}e^{\ii\omega_k t}-\frac{h_k}{\omega_k}b_k e^{-\ii\omega_k t} \right)}\,,\nn
&e^{\ii x \sigma^z}\sigma^{\pm}e^{-\ii x \sigma^z} =e^{\pm 2 \ii x} \sigma^{\pm}\,,
\end{align}
which leads by inserting the identity $U_{\rm p}^\dagger U_{\rm p}=\mathds{1}$ to:
\begin{align}
   \expval{\sigma^\pm} &= e^{\pm 2 \ii\left(\frac{\Delta}{2}t+\frac{\lambda}{\Omega}\sin(\Omega t)  \right)}\\
   &\quad\times\trace{U_{\rm p} e^{\pm2 \sum_k\left(\frac{h_k^*}{\omega_k}b_k^{\dagger}e^{\ii\omega_k t}-{\rm h.c.} \right)} U_{\rm p}^{\dagger}U_{\rm p} \sigma^{\pm}U_{\rm p}^{\dagger} \rho^0}\,.\nonumber
\end{align}
Next, the polaron transformation is applied to the system and bath parts separately:
\begin{align}
U_{\rm p} e^{\pm2 \sum\limits_k\frac{h_k^*}{\omega_k}b_k^{\dagger}e^{\ii\omega_k t}-{\rm h.c.}} U_{\rm p}^{\dagger}
   &= e^{\pm2 \sum\limits_k\frac{h_k^*}{\omega_k}\left(b_k^{\dagger}+\frac{h_k}{\omega_k}\sigma^z\right)e^{\ii\omega_k t}
   -{\rm h.c.} }\,,\nn
  U_{\rm p} \sigma^{\pm}U_{\rm p}^{\dagger} &=e^{\mp 2 \sum_k\left(\frac{h_k^*}{\omega_k}b_k^{\dagger}-\frac{h_k}{\omega_k}b_k \right)}   \sigma^{\pm}.
\end{align}
Separating system and bath parts and inserting the initial condition $\rho^0=\rho^0_S \otimes \bar{\rho}_B$ yields
\begin{align}
   \expval{\sigma^\pm} &= e^{\pm 2 \ii\left(\frac{\Delta}{2}t+\frac{\lambda}{\Omega}\sin(\Omega t)  \right)}\nn 
   &\qquad\times\trace{e^{\pm 4\ii \sum_k \frac{|h_k|^2}{\omega_k^2}\sin(\omega_k t) \sigma^z}\sigma^\pm \rho_S^0 } B_{\pm}(t)\,,\nn
  B_{\pm}(t)&={\rm Tr}\Big\{e^{\pm2 \sum_k\left(\frac{h_k^*}{\omega_k}b_k^{\dagger}e^{\ii\omega_k t}-\frac{h_k}{\omega_k}b_k e^{-\ii\omega_k t} \right)}\nn 
  &\qquad\times e^{\mp 2 \sum_k\left(\frac{h_k^*}{\omega_k}b_k^{\dagger}-\frac{h_k}{\omega_k}b_k \right)} \bar{\rho}_B\Big\}\,.
 \label{nebenrechnung}
\end{align}
The trace over the bath parts $B_{\pm}(t)$ gives: 
\begin{align}
    B_{\pm}(t)&=\exp\left\{\mp 4\ii \sum_k \frac{|h_k|^2}{\omega_k^2}\sin(\omega_k t) \right\}\\
    &\times\exp\left\{ -4\sum_k \frac{|h_k|^2}{\omega_k^2}[1-\cos(\omega_k t)] [ 1+2 n_B(\omega_k)]\right\}\,.\nonumber
\end{align}
Here, $n_B(\omega_k)=\frac{1}{e^{\beta \omega_k}-1}$ denotes the Bose distribution. Plugging the expression for $B_{\pm}(t)$ into Eq. (\ref{nebenrechnung}) and using 
\begin{align}
    {\rm Tr}\left\{e^{\pm 4\ii \sum_k \frac{|h_k|^2}{\omega_k^2}\sin(\omega_k t) \sigma^z}\sigma^\pm \rho_S^0 \right\}&= e^{\pm 4i \sum_k \frac{|h_k|^2}{\omega_k^2}\sin(\omega_k t)}\nn
    &\qquad\times \trace{\sigma^\pm \rho_S^0},
\end{align}
yields -- after inserting the spectral coupling density $\Gamma (\omega)=2 \pi \sum_k |h_k|^2 \delta (\omega-\omega_k)$ -- 
for the expectation value of $\expval{\sigma^+}=\rho_{10}$ Eq.~\eqref{EQ:pdp_exact} in the main text.


\section{Analytic fast-driving approximation}
\label{app fd}

We consider the simplified form of Eq.~\eqref{EQ:fast_driving} with $H_S^0=\frac{\Delta}{2} A$.
First, we move into an interaction picture with respect to the driving by performing the transformation $U_1(t)=e^{-\ii \frac{\lambda}{\Omega}\sin(\Omega t)C}$,
which yields after applying the rotating wave approximation
\begin{align}
   \tilde{H}_{\rm RWA}&=\tilde{A} \left(\frac{\Delta}{2}+\sum_k(h_k b_k+ h_k^*b_k^{\dagger})\right)+\sum_k\omega_k b_k^{\dagger} b_k\,,\nn
   \tilde{A}&=\frac{\Omega}{2 \pi}\int_0^{\frac{2 \pi}{\Omega}} U_1^\dagger(t) A U_1(t) dt\,.
\end{align}
This is the Hamiltonian of a simple pure dephasing system, see Eq. (\ref{EQ:pure_dephasing}), with time independent system Hamiltonian.

Applying the polaron transform~(\ref{polaron trafo}), system and bath part of the Hamiltonian can be decoupled and~(\ref{decoupled sb}) becomes
\begin{align}
    H_{\rm p}&=U_{\rm p}^\dagger \tilde{H}_{\rm RWA}U_{\rm p}\nn
    &= \frac{\Delta}{2} \tilde{A}- \sum_k \tfrac{|h_k|^2}{\omega_k}\tilde{A}^2+\sum_k h_k b_k^\dagger b_k\,,
\end{align}
which yields for the time evolution operator in this picture:
\begin{align}
    \tilde{U}(t)\approx e^{-\ii\left(\frac{\Delta}{2} \tilde{A}- \sum_k \tfrac{|h_k|^2}{\omega_k}\tilde{A}^2\right)t} e^{-\ii\sum_k h_k b_k^\dagger b_k t} .
\end{align}
In an anologous way as for the pure dephasing system, we can write the expectation value of an observable as:
\begin{align}
  \langle O_S  \rangle = \trace{ U_{\rm p} \tilde{U}^\dagger(t) U_{\rm p}^{\dagger}U_1^\dagger (O_S \otimes \f{1}_B)  U_1 U_{\rm p} \tilde{U}(t) U_{\rm p}^{\dagger}
\rho^0}.
\label{exp value fd}
\end{align}
For a two level system Eq. (\ref{tls fd}), we get
\begin{align}
    \tilde{A}= \mathcal{J}_0\left(\tfrac{2 \lambda}{\Omega}\right) \sigma^z,
\end{align}
with $\mathcal{J}_{n}(x)$ denoting the Bessel function of the first kind, which is defined as the solution of the differential equation $x^2 \mathcal{J}_{n}''(x)+x \mathcal{J}_{n}'(x)-(z^2+n^2)\mathcal{J}_n(x)=0$, eventually leading to~\eqref{EQ:fastdriving} in the main text.

\bibliography{references}

\end{document}